\documentclass[11pt, a4paper]{article}
\pdfoutput=1
\usepackage{jinstpub}
\usepackage{graphicx}
\newcommand{\gsim}      {\mbox{\raisebox{-0.4ex}{$\;\stackrel{>}{\scriptstyle \sim}\;$}}}

\title{The liquid-hydrogen absorber for MICE}

\collaboration{MICE collaboration}

\author[a]{V.~Bayliss,}
\author[a]{J.~Boehm,}
\author[a]{T.~Bradshaw,}
\author[a]{M.~Courthold,}
\author[a,\star]{S.~Harrison,}
\author[a]{M.~Hills,}
\author[b]{P.~Hodgson,}
\author[c]{S.~Ishimoto,}
\author[d]{A.~Kurup,}
\author[e]{W.~Lau,}
\author[d,a]{K.~Long,}
\author[a]{A.~Nichols,}
\author[f]{D.~Summers,}
\author[a,1]{M.~Tucker,}
\note{Corresponding author.}
\author[g]{P.~Warburton,}
\author[a,\dag]{S.~Watson,}
\author[h,i,j]{and C.~Whyte}

\affiliation[a]{STFC Rutherford Appleton Laboratory, Harwell Oxford, Didcot, OX11 0QX, UK}
\affiliation[b]{Department of Physics and Astronomy, University of Sheffield, Sheffield, S3 7RH, UK}
\affiliation[c]{KEK, Oho 1-1, Tskuba, Ibaraki 305-0801, Japan}
\affiliation[d]{Department of Physics, Blackett Laboratory, Imperial College London, London, SW7 2BB, UK}
\affiliation[e]{Particle Physics Department, The Denys Wilkinson Building, Keble Road, Oxford OX1 3RH, UK}
\affiliation[f]{Department of Physics, University of Mississippi-Oxford, University, MS 38677, USA}
\affiliation[g]{STFC Daresbury Laboratory, Daresbury, Cheshire, WA4 4AD, UK}
\affiliation[h]{SUPA, University of Glasgow, Glasgow, G12 8QQ, UK}
\affiliation[i]{Department of Physics, University of Strathclyde, Glasgow, G1 1XQ, UK}
\affiliation[j]{Cockroft Institute, Keckwick Lane, Daresbury, Warrington, WA4 4AD, UK}
\affiliation[\star]{Now at: AS Scientific Products Ltd, 2 Barton Lane, Abingdon, OX14 3NB, UK}
\affiliation[\dag]{Now at: Astronomy Technology Centre, Royal Observatory, Blackford Hill, Edinburgh, EH9 3HJ, UK}

\emailAdd{mark.tucker@stfc.ukri.org}

\abstract{
The Muon Ionization Cooling Experiment (MICE) has been built at the STFC Rutherford Appleton Laboratory to demonstrate the principle of muon beam phase-space reduction via ionization cooling.
Muon beam cooling will be required at a future proton-derived neutrino factory or muon collider.
Ionization cooling is achieved by passing the beam through an energy-absorbing material, such as liquid hydrogen, and then re-accelerating the beam using RF cavities.
This paper describes the hydrogen system constructed for MICE including: the liquid-hydrogen absorber, its associated cryogenic and gas systems, the control and monitoring system, and the necessary safety engineering.
The performance of the system in cool-down, liquefaction, and stable operation is also presented.
}


\begin{document}

\maketitle
\flushbottom

\section{Introduction}
\label{Sect:Intro}

Stored muon beams have been proposed as the source of neutrinos at a neutrino factory \cite{Geer:1997iz,Apollonio:2002en} and as the means to deliver multi-TeV lepton-antilepton collisions at a muon collider \cite{Neuffer:1994bt,Palmer:2014nza}.
In such facilities where the muon beam is produced from the decay of pions generated by a high-power proton beam striking a target, the muon beam occupies a large volume in phase space. 
To optimise the muon yield while maintaining a suitably small aperture in the muon-acceleration system requires that the muon beam be `cooled' (i.e., its phase-space volume reduced) prior to acceleration.
Recently, a muon-collider scheme based on the production of $\mu^{+} \mu^{-}$ pairs through the annihilation of positrons impinging on a target has been proposed \cite{Boscolo:2018tlu}.
These $\mu^{+} \mu^{-}$ pairs are created close to threshold which restricts the phase-space volume occupied by the muons and which could therefore produce beams with small emittance at energies greater than $\gsim$\,20\,GeV. 

A muon is short-lived, decaying with a lifetime of 2.2\,$\mu$s in its rest frame.
Therefore, beam manipulation at low energy ($\leq 1$\,GeV) must be carried out rapidly.
Four cooling techniques are in use at particle accelerators: synchrotron-radiation cooling \cite{2012acph.book.....L}; laser cooling \cite{PhysRevLett.64.2901,PhysRevLett.67.1238,doi:10.1063/1.329218}; stochastic cooling \cite{Marriner:2003mn}; and electron cooling \cite{1063-7869-43-5-R01}.
In each case, the time taken to cool the beam is long compared to the muon lifetime.
In contrast, the cooling time associated with ionization cooling, in which the energy of a muon beam is reduced as it passes through a material -- the absorber -- and is subsequently accelerated, is short enough to allow the muon beam to be cooled efficiently with acceptable decay losses.
Ionization cooling is therefore the technique by which it is proposed to reduce the muon-beam phase space in a proton-derived neutrino factory or muon
collider~\cite{cooling_methods,Neuffer:1983xya,Neuffer:1983jr,Stratakis2013,Stratakis2015}. 
This technique has never been demonstrated experimentally and such a demonstration is essential for the development of future high-brightness muon accelerators.

The international Muon Ionization Cooling Experiment (MICE) was designed \cite{MICEproposal:2003,MICE:2005zz} to perform a full demonstration of transverse ionization cooling.
Particle densities in the ionization-cooling channels that are conceived for the neutrino factory or muon collider are low enough for collective effects such as space charge to be neglected.
This allowed the MICE experiment to record muon trajectories one particle at a time.
The beam-cooling effect was produced by placing an energy-absorbing material in the bore of a superconducting solenoid that was used to focus and transport the muon beam.
The MICE collaboration has taken the data necessary to study the beam-cooling properties of lithium-hydride and liquid hydrogen.
This paper describes the design and construction of the containment vessel for the liquid hydrogen and the associated cryogenic and gas systems, together with a summary of the safety engineering and the control and monitoring system.
The performance of the system in cool-down, hydrogen liquefaction, and stable operation is also presented. 

\graphicspath{{Figures/}}

\section{The Muon Ionization Cooling Experiment}
\label{Sect:MICE}

\subsection{Beamline}
\label{Subsect:MICE:Beamline}

The muons for MICE were created from the decay of pions produced when a
target dipped into the circulating proton beam in
the ISIS synchrotron at the STFC Rutherford Appleton Laboratory
(RAL)~\cite{Booth:2012qz, Booth:2016lno}.
A beamline of nine quadrupoles, two dipoles, and a superconducting `decay
solenoid' collected each burst of particles (a `spill') and transported the 
momentum-selected particles to the experiment~\cite{Bogomilov:2012sr}.
The small fraction of pions that remained in the beam were identified, and subsequently rejected
during analysis, using the time-of-flight hodoscopes and Cherenkov
counters that were installed in the beamline upstream of the
experiment~\cite{Adams:2013lba}.

\begin{figure*}
  \begin{center}
    \includegraphics[width=\textwidth]{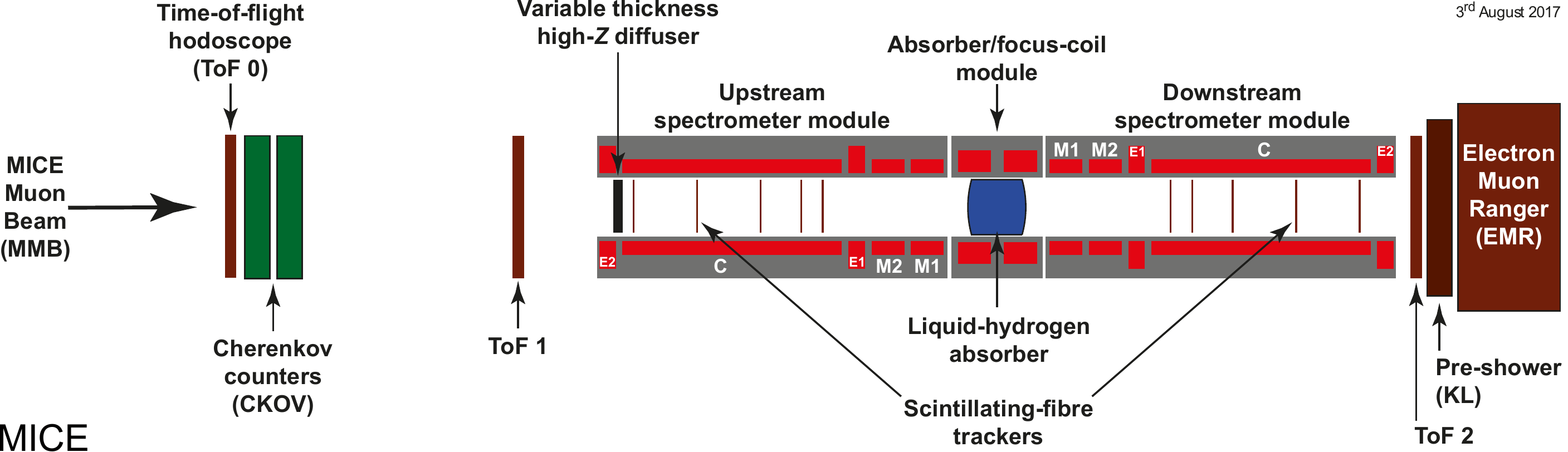}
  \end{center}
  \caption{
    Schematic of the experiment.
    The red rectangles represent the superconducting coils. 
    The individual coils in the spectrometer modules are labelled thus:
    E2, C, E1, M2, M1 in the upstream module, and M1, M2, E1, C, E2 in the downstream module.
    The various detectors (time-of-flight
    hodoscopes~\cite{Bertoni:2010by,MICE:Note:286:2010}, Cerenkov
    counters~\cite{Cremaldi:2009zj}, scintillating-fibre
    trackers~\cite{Ellis:2010bb}, KLOE Light (KL)
    calorimeter~\cite{Bogomilov:2012sr,Ambrosino2009239}, and
    electron muon ranger~\cite{Asfandiyarov:2016erh}) used to
    characterise the beam are also represented. 
  }
  \label{Fig:Overview}
\end{figure*}

The experiment contained an absorber/focus-coil (AFC) module sandwiched between two
spectrometer modules, as shown in figure \ref{Fig:Overview}.
The focus-coil (FC) module had two separate windings to produce
either aligned or opposed magnetic fields.
An absorber, such as lithium hydride (LiH) or liquid hydrogen (LH$_2$),
was placed at the centre of the FC module to `cool' the beam.

Each spectrometer module contained a long centre coil (C) with two end-correction
coils (E1, E2), and two `match' coils (M1, M2). 
The emittance was measured upstream and downstream of the cooling cell
using scintillating-fibre tracking detectors~\cite{Ellis:2010bb}
immersed in the uniform magnetic field provided by the coils (E1, C, E2).
The trackers were used to measure the trajectories of individual muons
before and after they had traversed the absorber. 
These reconstructed trajectories were combined with information from
instrumentation upstream and downstream of the spectrometer modules to
measure the muon-beam emittance at the upstream and downstream tracker
reference planes~\cite{Dobbs:2016ejn}.
A diffuser was installed at the inlet of the upstream spectrometer module to vary
the initial emittance of the beam~\cite{Blackmore:2011zz}. 
The instrumentation upstream and downstream of the spectrometer
modules enabled the selection --- for analysis --- of a pure sample of muons.
The `match' coils were used to match the beam optics between the
uniform-field region and the neighbouring FC.

\subsection{Safety considerations}
\label{SubSect:MICE:Safety}

The MICE hydrogen system was designed to store 22\,\textit{l} of liquid hydrogen
in an aluminium vessel at $\sim$20\,K and slightly above atmospheric pressure. 
Hydrogen/air mixtures are explosive between the lower explosive limit (LEL)
of 4\%  and the upper explosive limit (UEL) of 77\% 
hydrogen by volume.
Therefore, it was essential to prevent the ingress of air into the
hydrogen system and to prevent the escape of hydrogen into the
experimental hall.
Any venting of gas that may have been contaminated with hydrogen had to be
controlled such that it was purged into the atmosphere at a
concentration well below the LEL.

The philosophy adopted was to design a primary system to contain the 
hydrogen over a range of temperatures and pressures that
would exceed the range expected during operation.  
The room-temperature parts of this primary system were enclosed within a
secondary containment system which was continually flushed with
dry nitrogen gas. 
This ensured that any leak from this part of the primary system was
contained within an inert medium and continually flushed out to
atmosphere. 
The cooled part of the primary system was surrounded by an
insulating vacuum which
was continuously pumped and the exhaust released to atmosphere.
The entire hydrogen system, i.e. primary and secondary containment,
was engineered to be robust to two unrelated failures, whilst
maintaining safety in operation.
To meet these requirements a number of features were designed into the
system:
\begin {itemize}
  \item{\it Gas lines from bottle to point of use}: constructed
    from double-walled stainless-steel piping with continuous flow of dry
    nitrogen gas through the jacket.
    The internal tube and the jacket were both leak tested, and leak
    rates below $10^{-6}$\,mbar~\textit{l}/s were required; 
  \item{\it Process and control valves}: housed within a cabinet
    which was maintained at a pressure slightly below local ambient. 
    The cabinet was continually purged with dry nitrogen gas
    and vented through the roof;
  \item{\it Low-temperature circuit}: designed to ensure that
    there was no possibility of isolation of any volume of hydrogen
    due to freezing, thereby preventing any possible over-pressure and
    vessel-failure scenarios; and
  \item{\it Low-temperature circuit insulation vacuum}: its pumping system
    was vented via ATEX-rated pumps external to the experimental area and
    within a locked cabin to prevent personnel access during operation. 
    The vacuum-volume leak rate was verified to be below
    $10^{-8}$\,mbar~\textit{l}/s.
\end {itemize} 

All leak rates were measured using a helium leak detector when the system was warm and,
where relevant, verified once cold. Any air leaking into the vacuum of the secondary system
would have plated out on the cold surfaces and 
then been released if the system had warmed significantly 
for any reason. Such a warm-up would be expected if a significant leak of 
hydrogen into the insulating vacuum had occurred, potentially producing
an explosive mixture of hydrogen and air.
The maximum leak-rate into the vacuum was therefore chosen to restrict the volume of
plated gas accumulated over the operating period of the experiment to
a safe value.
Leak rates below $10^{-8}$\,mbar~\textit{l}/s would have resulted in a maximum
cryo-pumped air volume of $3 \times 10^{-4}$\,bar\,\textit{l} per year,
which was considered sufficiently low.  

\graphicspath{{Figures/}}

\section{Absorber vessel}
\label{Sect:AbsorberVessel}

\begin{figure}
  \begin{center}
    \includegraphics[width=0.95\textwidth]{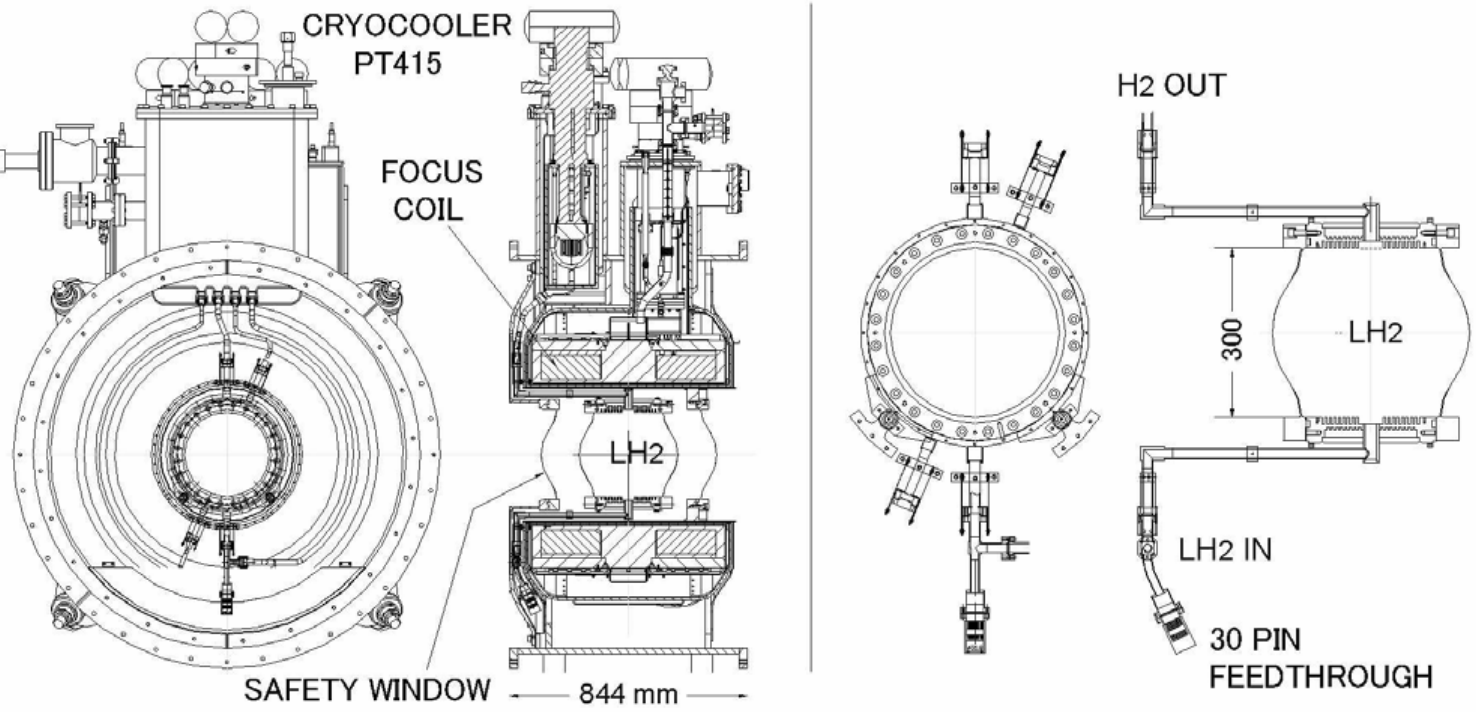}
  \end{center}
  \caption{
    Left panel: Drawing of the absorber/focus-coil (AFC) module showing the
    principal components.
    Right panel: detail of the liquid-hydrogen absorber vessel.
  }
  \label{Fig:AbsorberVessel:Diag}
\end{figure}

Drawings of the AFC module and the installed absorber vessel are shown in
figure~\ref{Fig:AbsorberVessel:Diag}.
The absorber vessel was set at the centre of the FC magnet coils.
The diameter of the warm bore was 470\,mm and its length
was 844\,mm. 
The diameter of the flange by which the FC was connected mechanically
to the spectrometer module was 1,514\,mm. 

\subsection{Design considerations}
\label{SubSect:AbsorberVessel:Design}

As a muon beam passes through material, some of the kinetic energy
of the muons is lost through ionization of the material.
This process results in a reduction of the normalised
transverse emittance and the beam is said to be cooled.
Muons will also undergo multiple Coulomb scattering which
increases the divergence of the beam, thereby
increasing the normalised transverse emittance and heating the beam.

Ionization-energy loss is characterised by $\frac{dE}{dx}$, where $E$
is the muon energy and $x$ is the distance travelled within the
absorber.
Multiple Coulomb scattering is characterised by the radiation length,
$X_0$.
For liquid hydrogen, $\frac{dE}{dx} \sim 0.03$\,MeV/mm and $X_0 \sim
8905$\,mm~\cite{Tanabashi:2018xxx}.
The absorber vessel was manufactured using aluminium for which
$\frac{dE}{dx} \sim 0.4$\,MeV/mm and $X_0\sim
90$\,mm~\cite{Tanabashi:2018xxx}.
To maximise the cooling effect from energy loss in liquid hydrogen,
while minimising the heating effect from multiple Coulomb scattering
in the aluminium windows, these windows were required to be as thin as possible.
Safety considerations, as described in section~\ref{SubSect:MICE:Safety},
required a secondary containment system.
Therefore, the absorber vessel was situated in an evacuated space within
two more thin aluminium safety windows, so the muon beam had to traverse four windows,
as shown in the left panel in figure~\ref{Fig:AbsorberVessel:Diag}.

\subsection{Absorber vessel body}
\label{SubSect:AbsorberVessel:Body}

The absorber vessel comprised a cylindrical aluminium body sealed with
two thin aluminium end windows, as shown in the right panel of
figure~\ref{Fig:AbsorberVessel:Diag}.
The absorber vessel was specified to contain 22\,\textit{l} of liquid, so
the body had an inner diameter of 300\,mm and a length between its end
flanges of 230\,mm.  
The length along the central axis between the two domes of the thin aluminium end
windows was 350\,mm.
The body contained an annular cooling channel within its walls
that could act as a heat exchanger. 
This channel was designed to allow the possibility of cooling the
vessel body directly using liquid nitrogen, or even liquid helium. 
However, it was found that this cooling was not necessary because the absorber vessel
cooled sufficiently quickly with cold gas from the condenser,
as described in section~\ref{Sect:Subsect:Helium}. 
Small indium-sealed flanges connected the aluminium pipes from
the absorber vessel to the stainless-steel pipes from the condenser. 

\begin{figure}
  \begin{center}
    \includegraphics[width=0.95\textwidth]{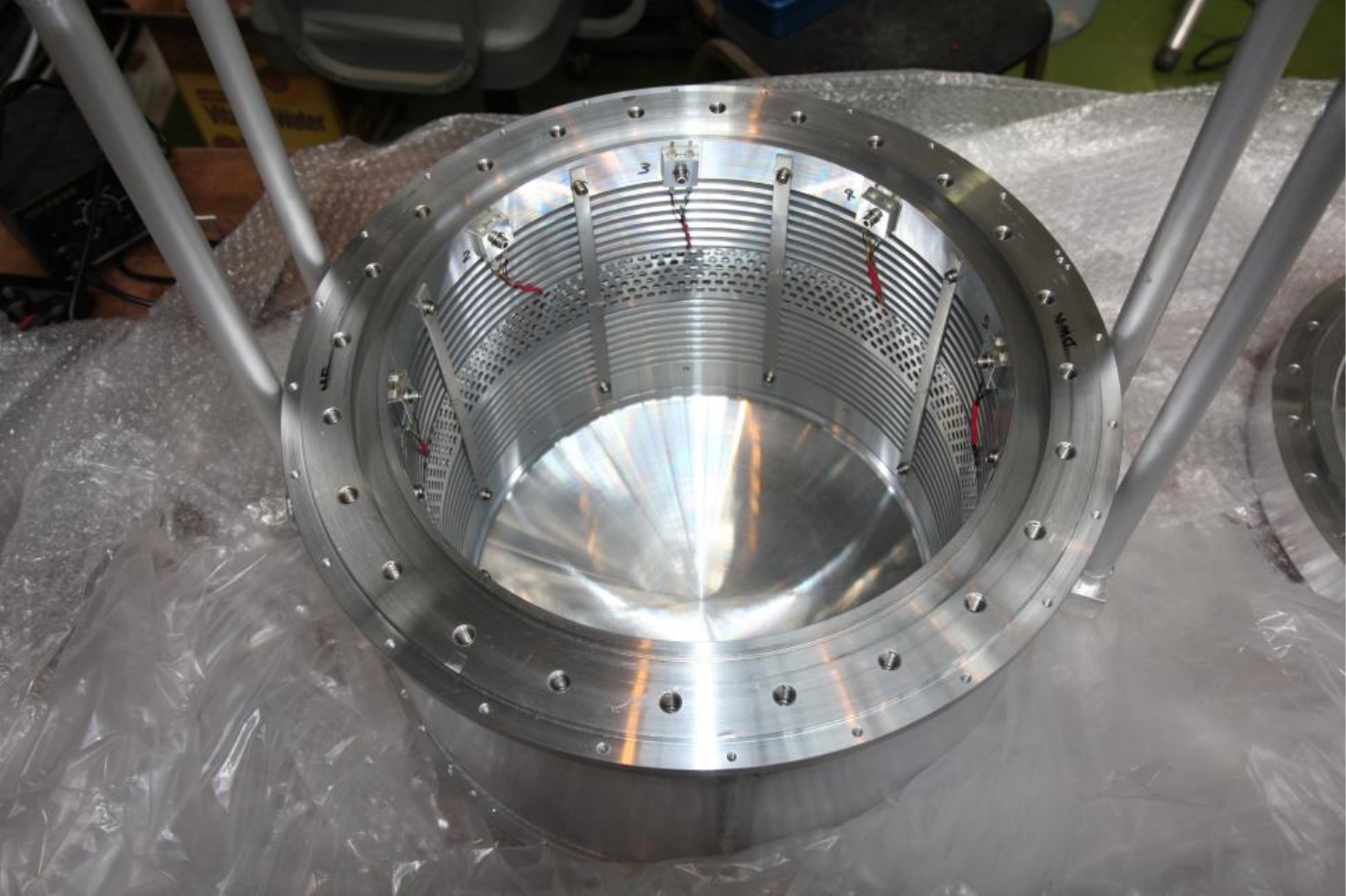}
  \end{center}
  \caption{
    Photograph of the absorber vessel body.
  }
  \label{Fig:AbsorberVessel:BdyPhoto}
\end{figure}

Figure \ref{Fig:AbsorberVessel:BdyPhoto} shows a photograph of the inside of the
absorber vessel body.   
The two flanged windows were sealed to the end flanges of this
body using indium contained in grooves. 
The heat exchanger fins and five pairs of thermometers (LakeShore Cernox 1050-SD)
are visible in this photograph. 
These five thermometer pairs were inside the vessel at locations spaced by 45$^{\circ}$ around
the circumference and were monitored with a LakeShore 218S.
Each pair monitored the presence of liquid hydrogen at that position;
one of these Cernox sensors was operated with a small current as a
thermometer, and the other was occasionally heated by a pulse of larger current.
The difference between the two measured temperatures was small when these sensors
were in liquid due to good cooling efficiency, but the difference was
larger when these sensors were in gas since heat transport through the
gas is worse than in the liquid.
The sensor wires were extracted to vacuum part-way along the
liquid-hydrogen inlet pipe at a 30-pin hermetic feed-through, as
shown in figure~\ref{Fig:AbsorberVessel:Diag}.
Signals from each sensor were carried on two wires inside the absorber
vessel, between the sensor and the feed-through, and by four wires in
the vacuum outside the vessel.
Two Cernox thermometers and two heaters (LakeShore HR-25-100) were
mounted externally on each end flange.
Two additional Cernox
thermometers were mounted externally on the hydrogen inlet and outlet lines.
These thermometers were exposed to vacuum and thermal radiation so the thermometry
here was less accurate than that inside the absorber vessel, but gave indications
of the flow of cooling gas in the circuit. 
To minimise heat input from contact with the magnet bore, the absorber vessel was
mounted on glass-epoxy (G10) supports of low thermal conductivity.
To minimise radiative heat input, multilayer
insulation (MLI) was wrapped around the absorber vessel and all
the low-temperature pipework. 
The number of layers of MLI over the end windows was first entered
into a Monte Carlo program to check that the scattering of muons by
the MLI was insignificant compared to that of the windows, before the
vessel was integrated into the system and cooled.

\subsection{Windows}
\label{SubSect:AbsorberVessel:Wndws}

\begin{figure}
  \centerline{\includegraphics*[width=0.99\textwidth]{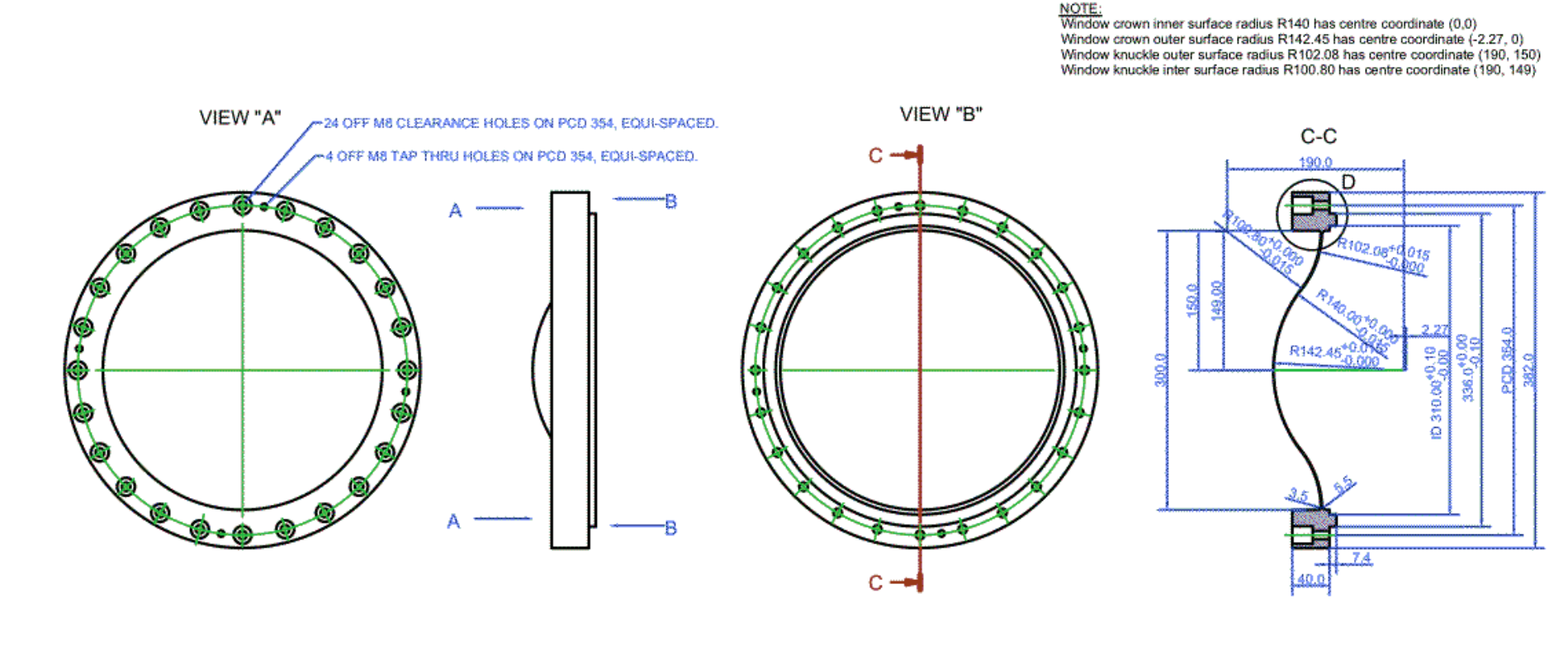}}
  \caption{
    Aluminium absorber vessel window with a central thickness of 180 $\mu$m
    for the containment of liquid hydrogen. 
    Both types of safety windows were similar to the vessel window,
  but had a central design thickness of 210\,$\mu$m.
  } 
  \label{AbsorberWindow}
\end{figure}

The liquid hydrogen was contained between two aluminium
windows, each having a thickness of 180\,$\mu$m at the centre and
increasing to 360\,$\mu$m near the outer flange. 
Aluminium safety windows, each with a central thickness of 210\,$\mu$m,
enclosed the absorber vessel in the magnet bore.  
Thin aluminium was chosen to minimise multiple scattering.
Thinner windows lead to less scattering
and more muon-beam cooling.
Although a MICE window with a central thickness of 125\,$\mu$m had
successfully been machined using alloy 6061-T651, it  would not withstand
enough pressure. 
The pressure in the absorber vessel reached 1500 mbar during typical operations.
The aluminium alloy we chose to use (6061-T651) was assayed to contain 
 0.61\% silicon, 
 0.26\% iron,
 0.25\% copper,
 0.02\% manganese,
 1.02\% magnesium,
 0.20\% chromium,
 0.01\% zinc, 
 0.05\% titanium,
 0.01\% zirconium, 
 0.15\% maximum other material, 
and at least 97.42\% aluminium (all measured by weight). 
The yield strength was measured at room temperature to be
39,900$\pm$700\,psi (275$\pm$5\,MPa),   
although this would be greater at 20\,K. 
A drawing of a MICE absorber vessel window is shown in
figure~\ref{AbsorberWindow}. 
The double-bend geometry increases the burst strength.

\subsubsection{Window manufacture}

A CNC Fadal 5020A vertical machining centre and a CNC Romi lathe with
a 27 inch swing were used to machine the windows from a solid block of
aluminium alloy. 
Precision backing plates supported the windows during this process.
Each window was machined to a 2000\,$\mu$m central thickness, and then
measured with the micrometer jig shown in figure~\ref{micrometers}. 
The window was then returned to the lathe for final machining while the
lathe still had the positions stored in its memory.
Clear plastic cases were fabricated to protect the windows from damage
in transit, while still allowing visual inspection.
Finished windows can be seen in figures~\ref{LBL_CMM}
and~\ref{beta_rays}. 

\subsubsection{Window thickness measurement}

\begin{figure}
  \begin{minipage}{0.46\textwidth}
    \centerline{\includegraphics*[width=0.9\textwidth]{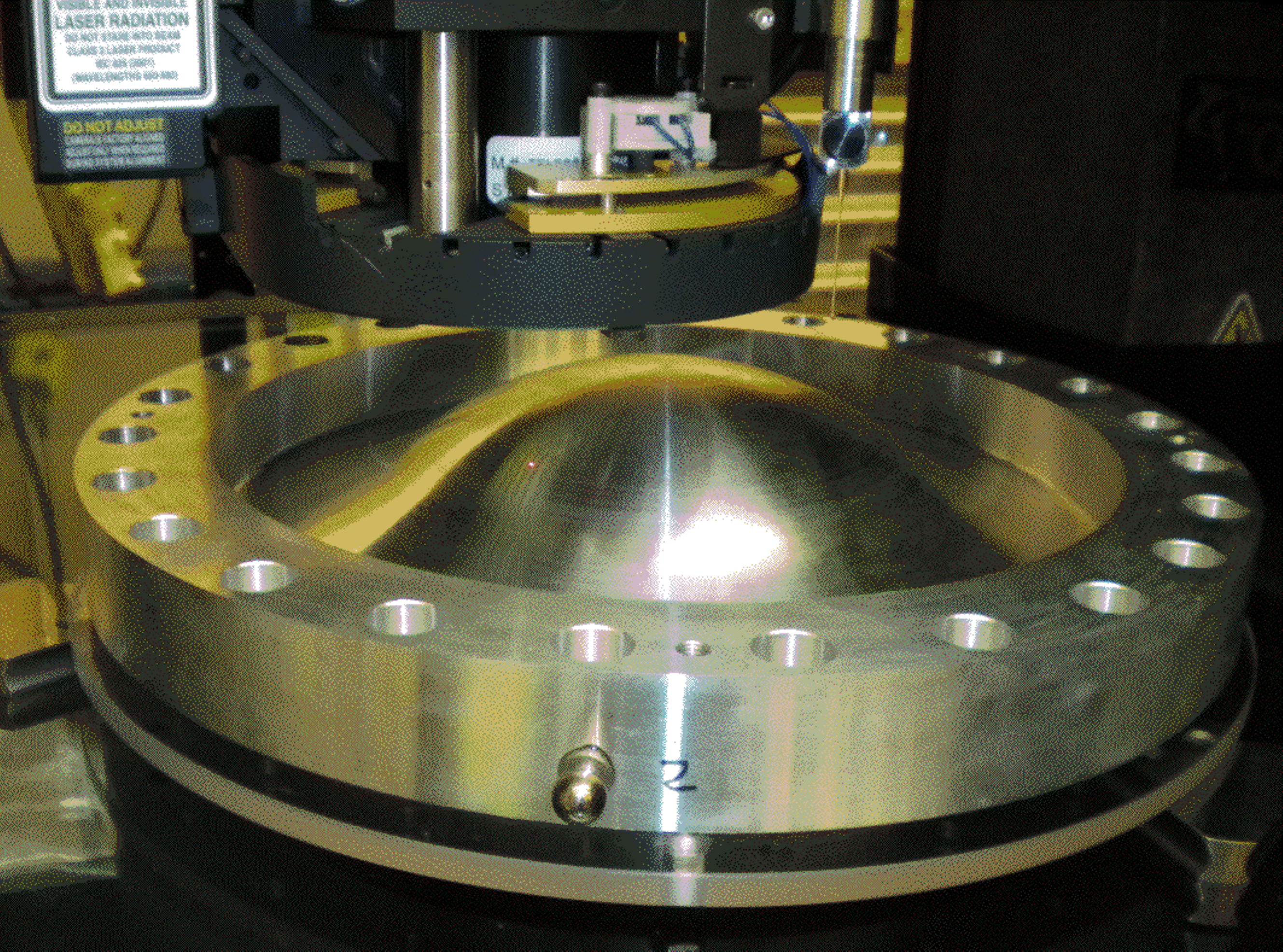}}
    \caption{
      The View Precis 3000 Optical CMM measured the surface profile of
      each window, one side at a time.
    }
    \label{LBL_CMM}
  \end{minipage}\hfill%
  \begin{minipage}{0.46\textwidth}
    \centerline{\includegraphics*[width=0.85\textwidth]{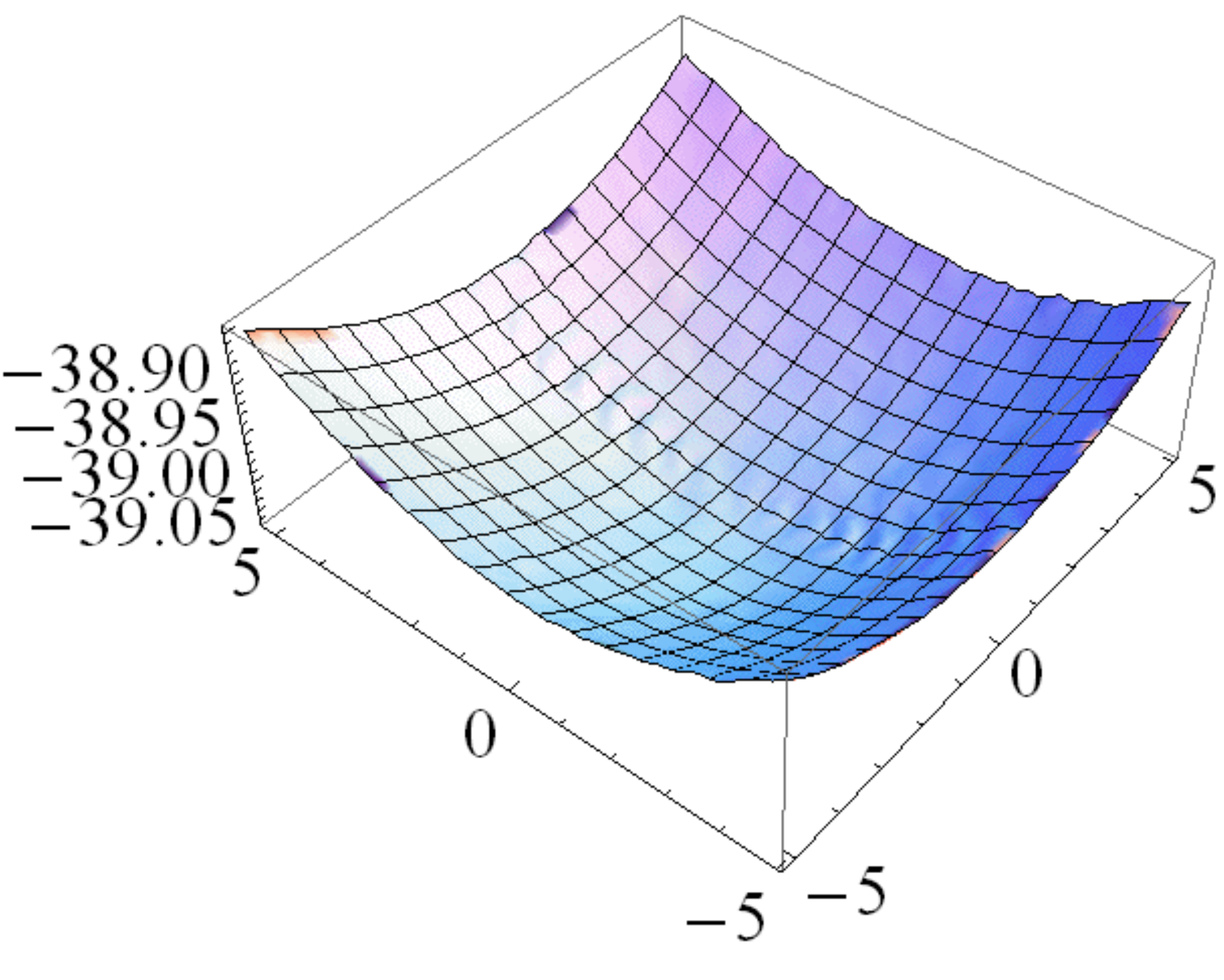}}
    \caption{
      Result of the CMM measurement of one side of one window. All the axes are
      labelled in units of millimetres.
    }
    \label{LBL_CMM2}
  \end{minipage}
\end{figure}

The thicknesses of three different types of finished windows
(one absorber and two safety) were measured
with the View Precis 3000 Optical Co-ordinates Measurement Machine
(CMM) shown in figure~\ref{LBL_CMM}.
The complete surface profile of a window was measured with the laser
on one side, and then the window was turned over to measure the other
side.  
The difference between the surface profiles of both sides of the
window gave the thickness.  
Three tuning balls were glued to the window to establish the reference
coordinate system; key to getting a good measurement was to establish
the same reference coordinate system for both sides of each window.
Some results of the measurements are
shown in table~\ref{tab:summary} and figure~\ref{LBL_CMM2}.
For some of the windows, the thickness measurement was checked by
scanning only the small area around the window centre with a very
dense meshing. 
This gave a more accurate measurement of the thickness at the window
centre.

\begin{table}
  \caption{
    Results of measuring the central thickness of the three types of
    windows with the View Precis 3000 Optical CMM shown in
    figure~\ref{LBL_CMM}.
    The windows actually used in MICE were numbers 002, 003, 009, and
    014.
  }
  \label{tab:summary}
  \begin{center}
    \begin{tabular}{|c c c  c c|}
    \hline
Window & Window & Central Thickness & Central Thickness & Note    \rule{0pt}{14pt} \\
 \#  &    Type      & Measured ($\mu$m)              & Design ($\mu$m) &  \\
\hline
001 & Absorber& & 180 & \\
002 & Absorber& $174\pm5$ &180&\\
003 & Absorber& $184\pm2$ & 180&\\
004 & Absorber&& 180&\\
005 & Absorber& $176\pm6$ & 180 &\\
006 & Safety I& $222\pm6$ &210 & flaw at centre\\
007 & Safety I& & 210&  flaw at centre\\
008 & Safety II& $233\pm$5 &210 &\\
009 & Safety II& $230\pm$9 &210 &\\
010 & Absorber&  & 180&\\
011 & Absorber&  & 180&\\
012 & Safety I& $197\pm 7$ & 210  & \\
013& Safety I & & 210  &\\
014 & Safety I & $197\pm8$ & 210 &\\
    \hline
    \end{tabular}
  \end{center}
\end{table}

\begin{figure}
  \begin{minipage}{0.46\textwidth}
    \centerline{\includegraphics*[height=0.73\linewidth]{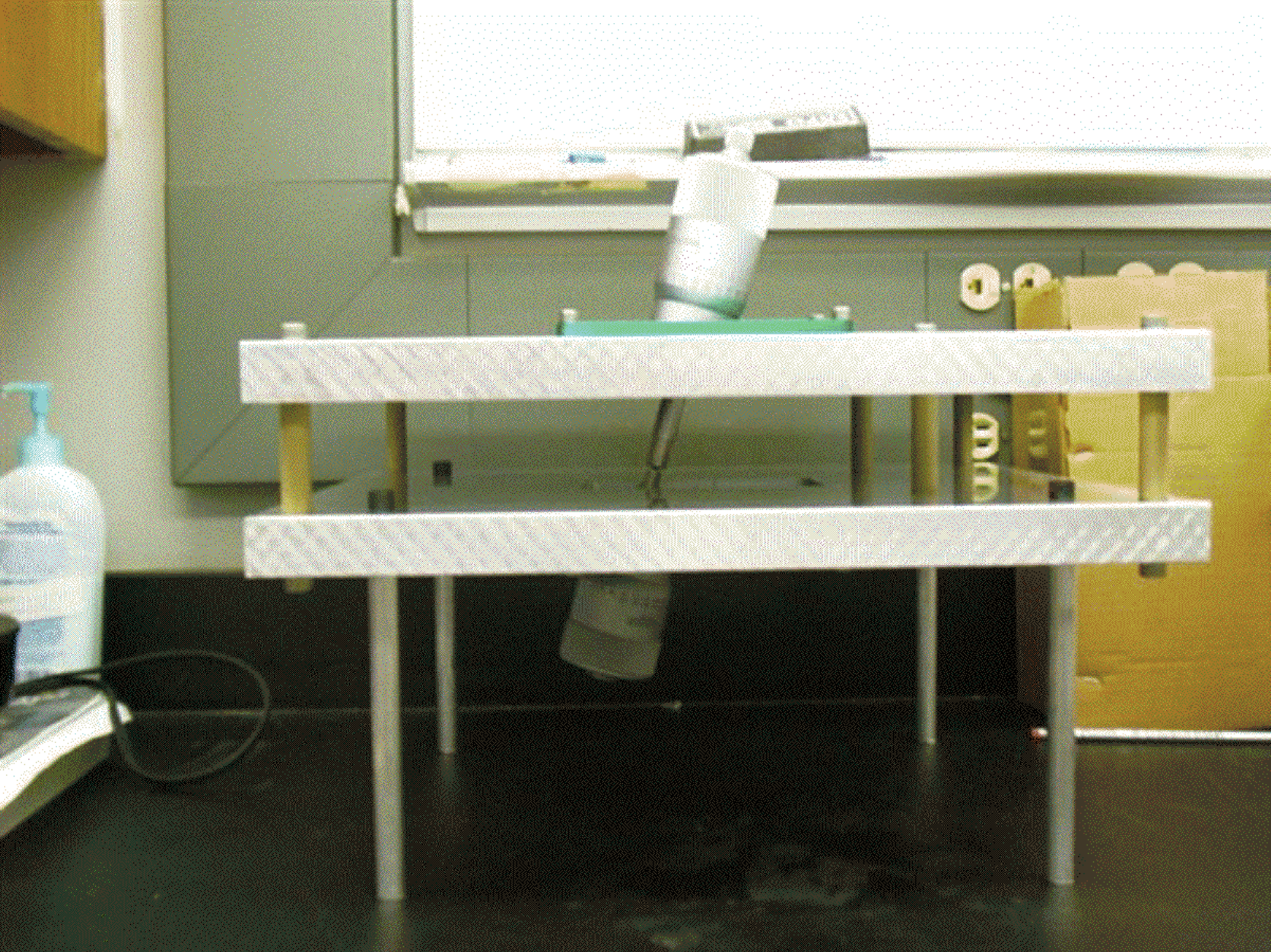}}
    \caption{
      Jig for measuring window thickness at the centre and at 15$^{\circ}$
      from the peak of the dome with a pair of
      Starrett T465 micrometers accurate to 3 microns.
    } 
    \label{micrometers}
  \end{minipage}\hfill%
  \begin{minipage}{0.46\textwidth}
    \centerline{\includegraphics*[height=0.73\linewidth]{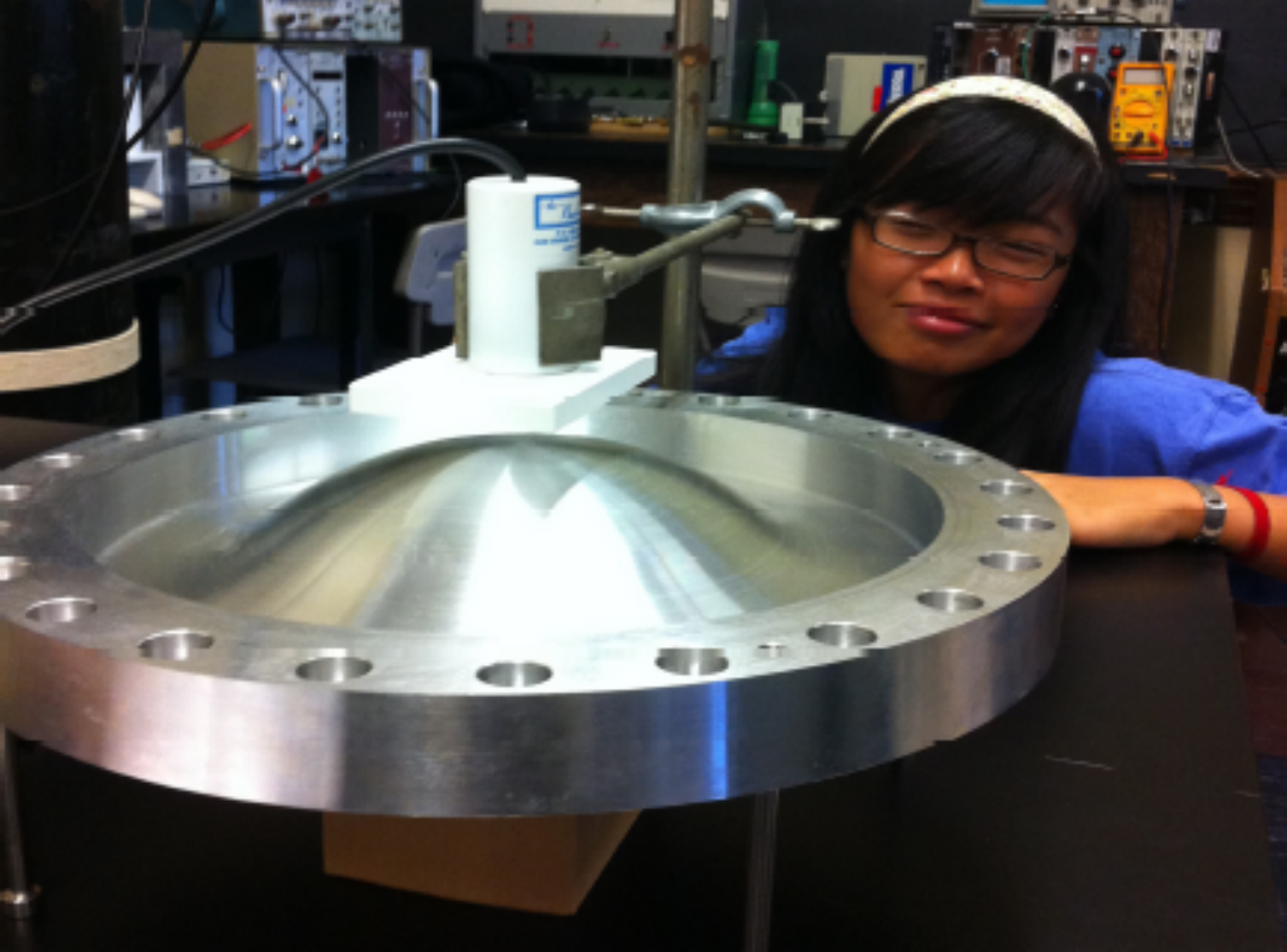}}
    \caption{
      $^{90}$Sr and $^{204}$Tl beta sources and a Geiger tube were used to
      check the central thicknesses of windows.
    }
    \label{beta_rays}
  \end{minipage}
\end{figure}

Low energy electrons are strongly attenuated by modest thicknesses of aluminium. 
Two different beta sources, $^{90}$Sr and $^{204}$Tl, were used to measure the thickness of a MICE
window. The source and detector (Geiger tube) were on opposite sides
of the window so there was no need to 
move the window during this process, as was required with the laser CMM. 
The attenuation of electrons in a thin sheet of material of
thickness $x$ was described using the equation:
\begin{equation}
  R = A \, e^{\,\alpha x} + B\, e^{\,\beta x} + C.
\end{equation}
The apparatus was optimised to measure the central window thickness by
choosing beta sources with electron energies that have a half-range of
about 180\,$\mu$m in aluminium.
Due to electron scattering, the result can be sensitive to apparatus
geometry so a careful calibration was performed using aluminium sheets
of known thickness with
counts being accumulated for 10\,minutes per sheet.
The central thickness of the absorber window in figure~\ref{beta_rays}
was measured to be $178 \pm 6\,{\rm (stat)} \pm 4\,{\rm (fit)}$\,$\mu$m.

\begin{figure}
  \begin{minipage}{0.46\textwidth}
    \centerline{\includegraphics*[height=0.76\textwidth]{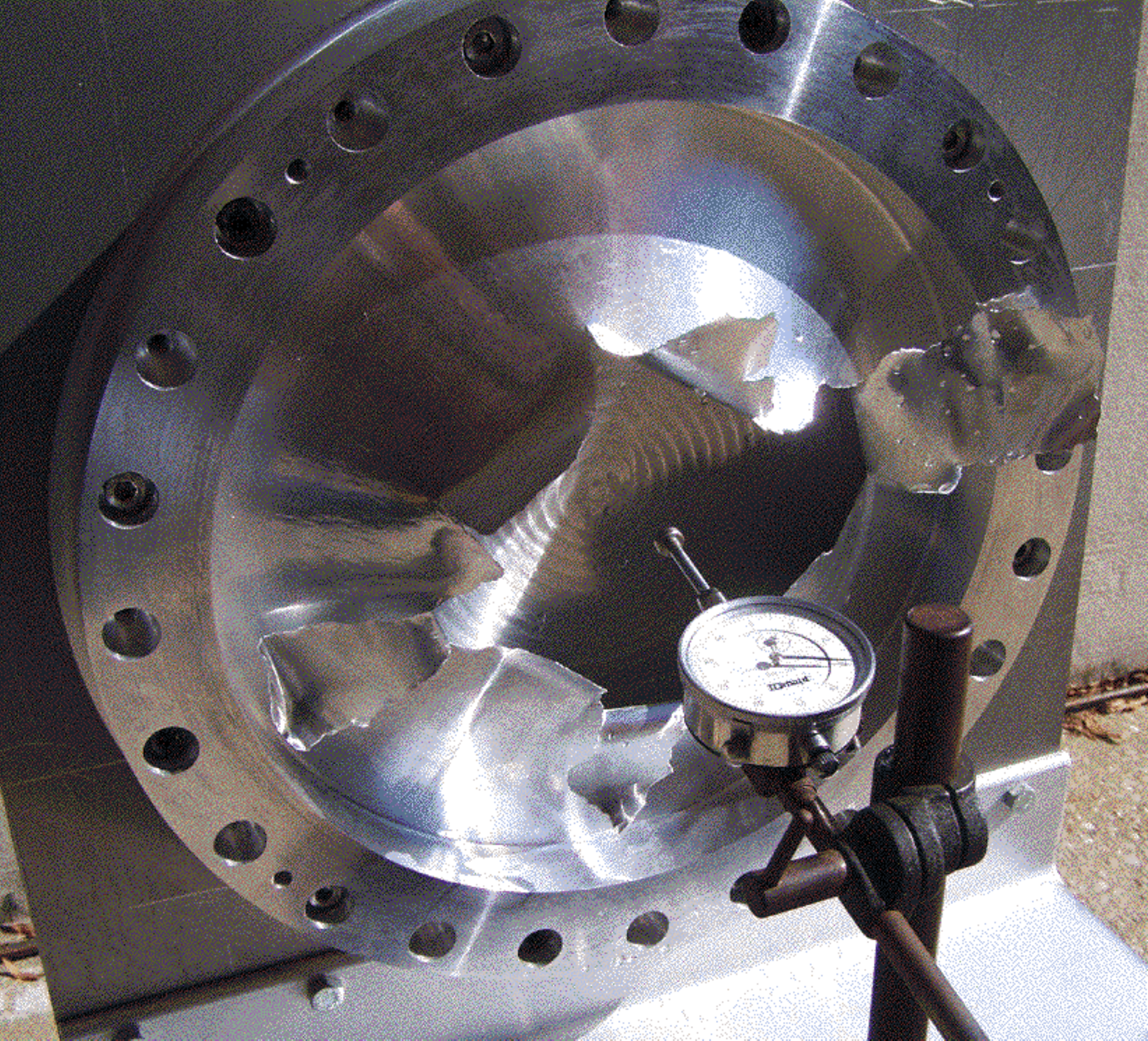}}
    \caption{
      This absorber vessel window burst when pressurised with water.
    }  
    \label{Burst_Absorber}
  \end{minipage} \hfill
  \begin{minipage}{0.46\textwidth}
    \centerline{\includegraphics*[height=0.75\textwidth]{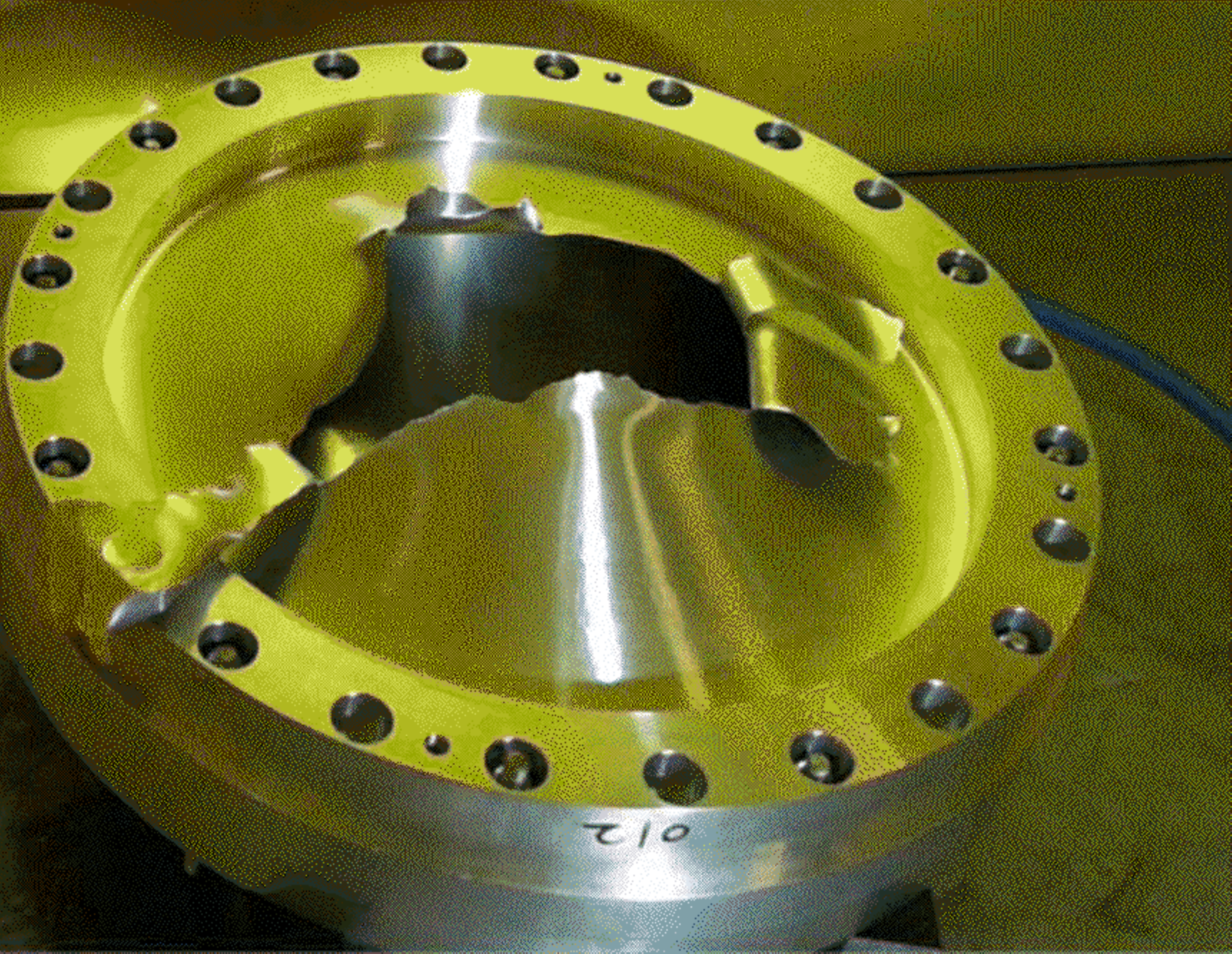}}
    \caption{
      This Type I safety window (number 012 in Table \ref{tab:summary})
      burst when slowly pressurised with nitrogen gas.
    } 
    \label{Burst_Safety}
  \end{minipage}
\end{figure}

\subsubsection{Window burst tests}

The windows were designed to withstand internal pressures from vacuum
($-$1.0\,bar) to a minimum of 96\,psig. (6.6\,bar).
Two absorber windows (figure~\ref{Burst_Absorber}) and one safety
window (figure~\ref{Burst_Safety}) had been destructively
burst-tested at room temperature.
A dial indicator showed the deflection up to the moment of burst.
The absorber windows burst at 8.27\,bar and 8.41\,bar, respectively,
and the Type I safety window burst at 7.67 bar.

\graphicspath{{Figures/}}

\section{Condensing unit}
\label{Sect:H2Condens}

\begin{figure}
  \begin{center}
    \includegraphics[width=0.85\textwidth]{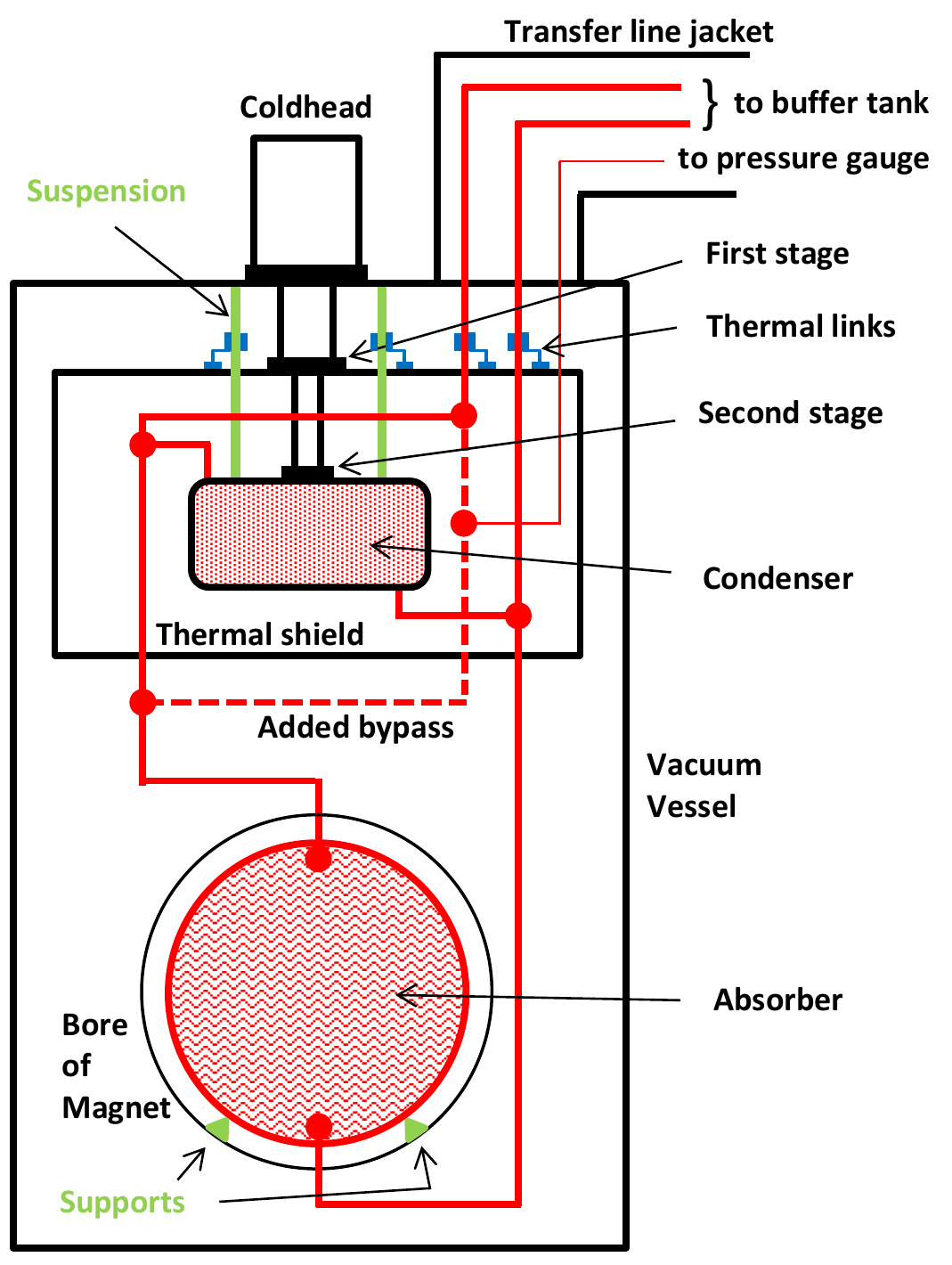}
  \end{center}
  \caption{
    Schematic of the condensing system. The bypass loop added as an
    extra precaution against blockage by solidified hydrogen is shown as a dashed line.
  }
  \label{Fig:H2Condens:Schema}
\end{figure}

\begin{figure}
  \begin{center}
    \includegraphics[width=0.95\textwidth]{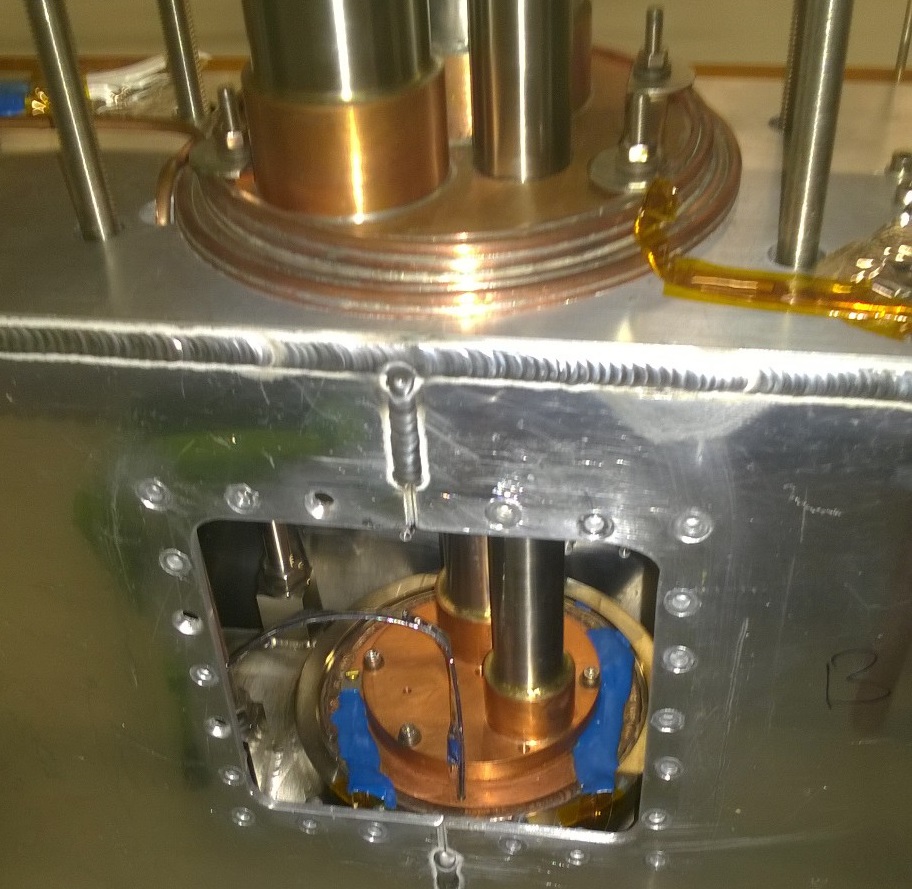}
  \end{center}
  \caption{
    Through the access window in the thermal shield can be seen the
    bolted thermal connection between the second stage of the coldhead
    and the condenser.
    The condenser is mostly hidden inside the welded aluminium thermal
    shield that was connected (at the top of the picture) to the first
    stage of the coldhead. The window was blanked with an aluminium
    plate before the shield was fully enclosed in blankets of
    super-insulation for the cooldown. 
    The blue stycast bonded the wires of the two thermometers to the
    copper plate for good heat-sinking.
    At the front of the copper plate, the vertical wire bundle wrapped
    in super-insulation connected to the heater which had been inserted
    into a drilled hole.
    This control heater prevented the condenser from getting too cold
    and thus prevented freezing of the hydrogen.
  }
  \label{Fig:H2Condens:PotPic}
\end{figure}

The condenser was situated above the absorber vessel and both were inside the
thermally-insulating vacuum of the safety volume inside the cryostat,
as shown in figure~\ref{Fig:H2Condens:Schema}. 
The condenser was cooled through direct contact with the
second stage of a coldhead
in order to cool, and ultimately liquefy, the incoming
gas to fill the absorber vessel with liquid.
The condenser was suspended from the underside of the top-plate of the cryostat by
four stainless-steel rods, with couplings that allowed for differential contraction
of the rods and the coldhead. 
These rods were heat-sunk to the thermal shield to reduce the 
thermal conduction to the condenser along these four rods to $\sim 0.15$\,W. 
Inside the condenser was an array of copper fins to facilitate the
removal of heat from the gas. 
These fins were part of a large copper block, fitted into the curved
side of the stainless-steel condenser, to which the second stage of the coldhead was
firmly connected,
as can be seen in figure~\ref{Fig:H2Condens:PotPic}.

\begin{figure}
  \begin{center}
    \includegraphics[width=0.95\textwidth]{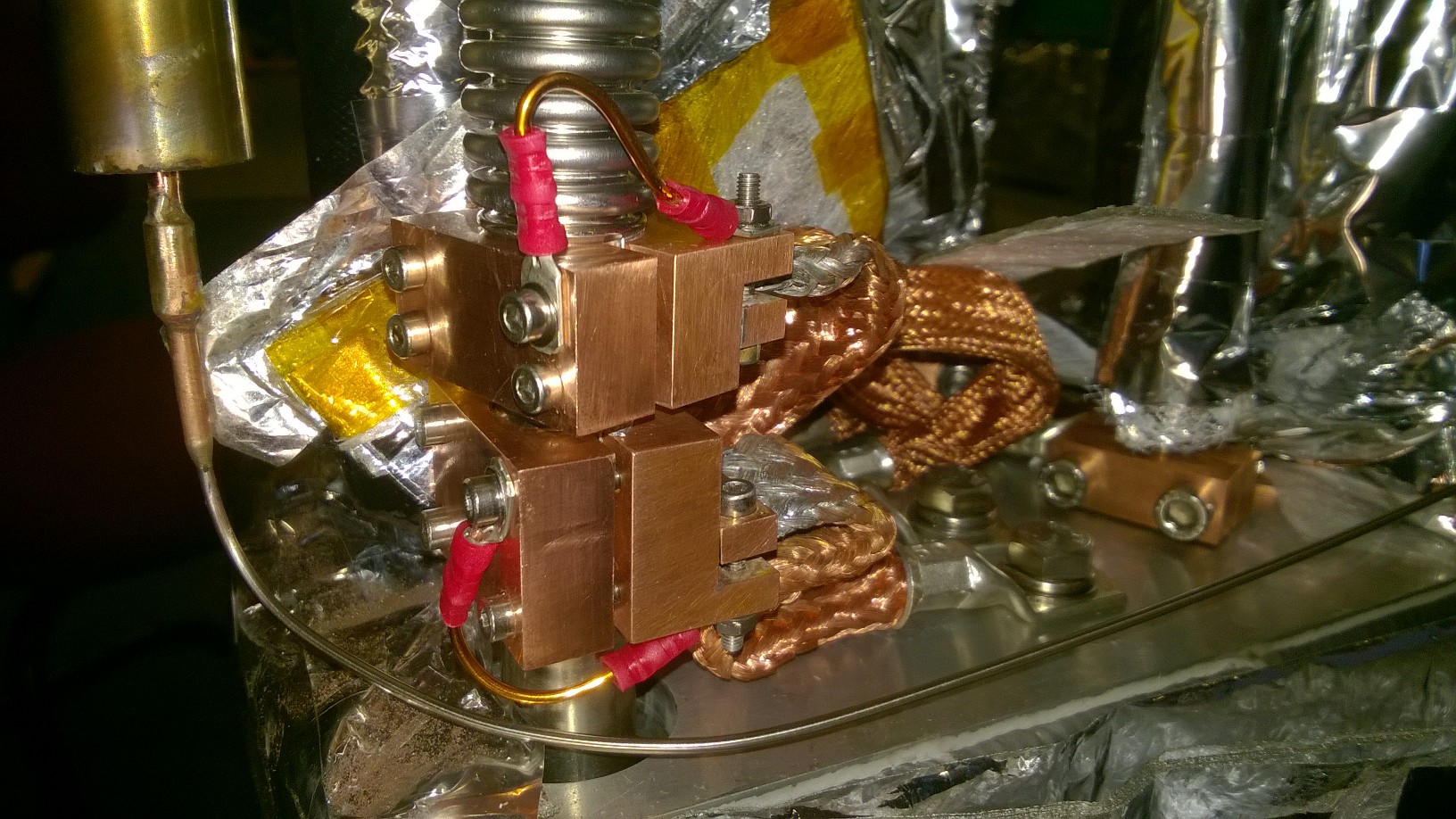}
  \end{center}
  \caption{
    Photograph showing some of the copper heat-links between the
    gas-filled tubes and the thermal shield before being covered with
    three blankets of MLI.
  }
  \label{Fig:H2Condens:ThermShldPic}
\end{figure}

The condenser was cooled by the second stage of the two-stage coldhead
(Cryomech PT415 pulse tube); this second stage has a nominal
cooling capacity of 1.5\,W at 4.2\,K and base temperature of 2.8\,K. 
The first stage of the coldhead cooled the thermal shield that
surrounded the condenser to about 45\,K, as well as the incoming gas via thermal
links between this shield and the gas-filled tubes, as  shown in
figure~\ref{Fig:H2Condens:ThermShldPic}.
The thermal shield was covered with three blankets of multi-layer
insulation (MLI); each blanket comprised ten layers of Mylar coated
with reflective aluminium film, the layers separated by polyester netting.
The 15\,mm bore stainless steel pipework between the condenser and the
absorber vessel was covered with two blankets of MLI. 
The absorber vessel, including its thin windows, was covered with four
blankets of MLI. 
For lack of space, it was not possible to install a cooled thermal
shield around the pipework and the absorber vessel; 
this would have improved the rates of cooling and liquefaction. 
The weight of the pipework near the bottom of the absorber vessel was
suspended by six strands of 0.5\,mm diameter nickel alloy wire of length
450\,mm.
These low-thermal-conductivity wires made a negligible contribution of about
0.01\,W to the heat-load.

\begin{figure}
  \begin{center}
    \includegraphics[width=0.95\textwidth]{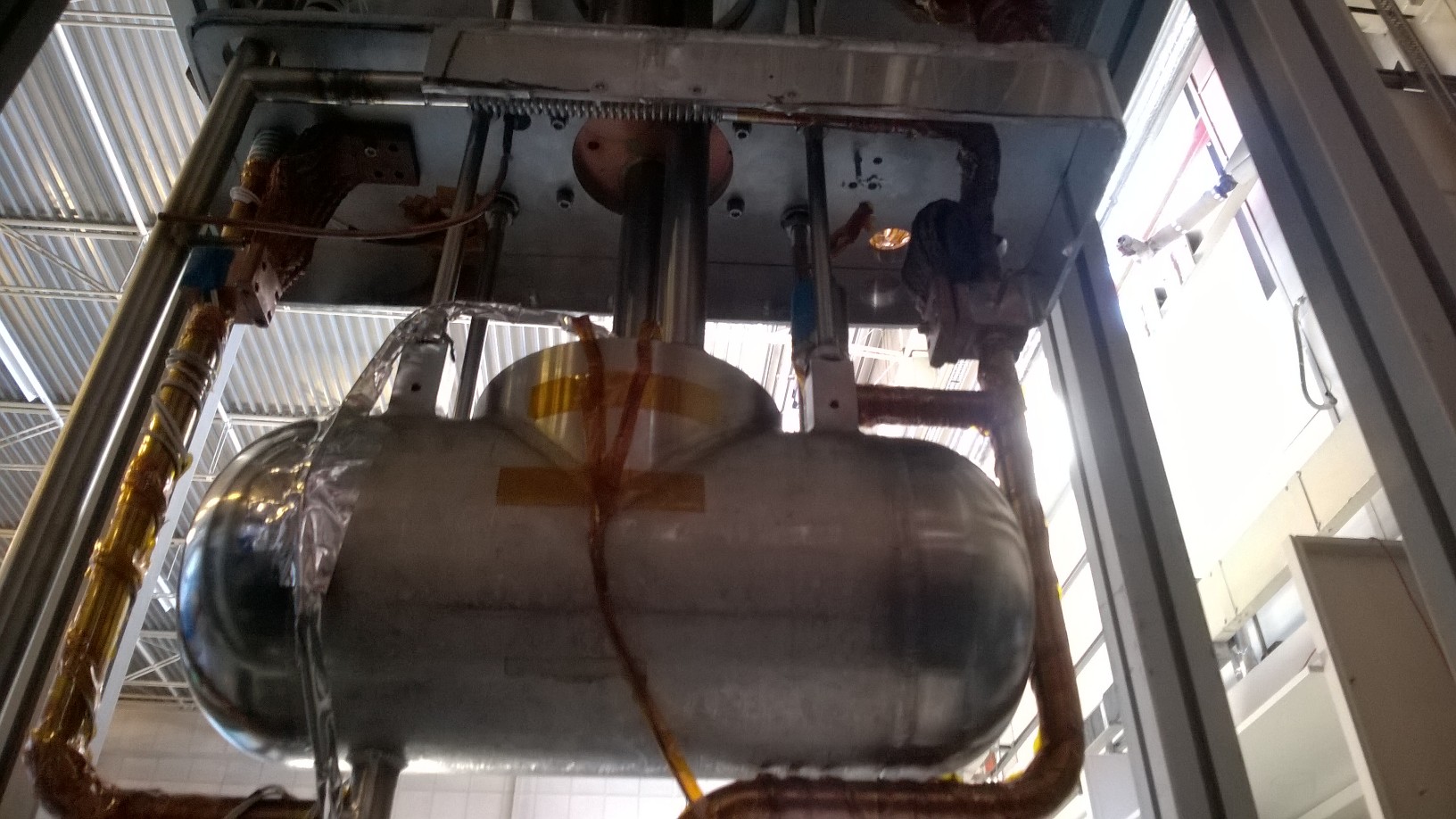}
  \end{center}
  \caption{
    Photograph showing the condenser with the bypass loop
    added. The aluminium thermal shield had been cut away for this
    modification and was re-welded for the following runs.
  }
  \label{Fig:H2Condens:ByPassPic}
\end{figure}

To avoid the possibility of blocking the pipework due to
the formation of hydrogen-ice (under fault conditions which could cause temperatures lower than 
$\sim 14\,$K in the condenser), a bypass loop was included as
shown in figures~\ref{Fig:H2Condens:Schema} and~\ref{Fig:H2Condens:ByPassPic}.
The lower end of the bypass was at the (never-freezing) boil-off side
of the absorber pipework and the upper end was at the (never-freezing)
first stage of the coldhead, the temperature of which could never go
below $\sim 35$\,K.
Thus, the bypass line always provided an ice-free path from the absorber vessel to the pressure
relief valves.
This bypass line had an insignificant effect on the cooling
efficiency.

\graphicspath{{Figures/}}

\section{Gas system}
\label{Sect:GasPanel}

\subsection{Overview of gas flows}
\label{SubSect:GasPanel:OvrvwGasFlow}

\begin{figure}
  \begin{center}
    \includegraphics[width=0.85\textwidth]{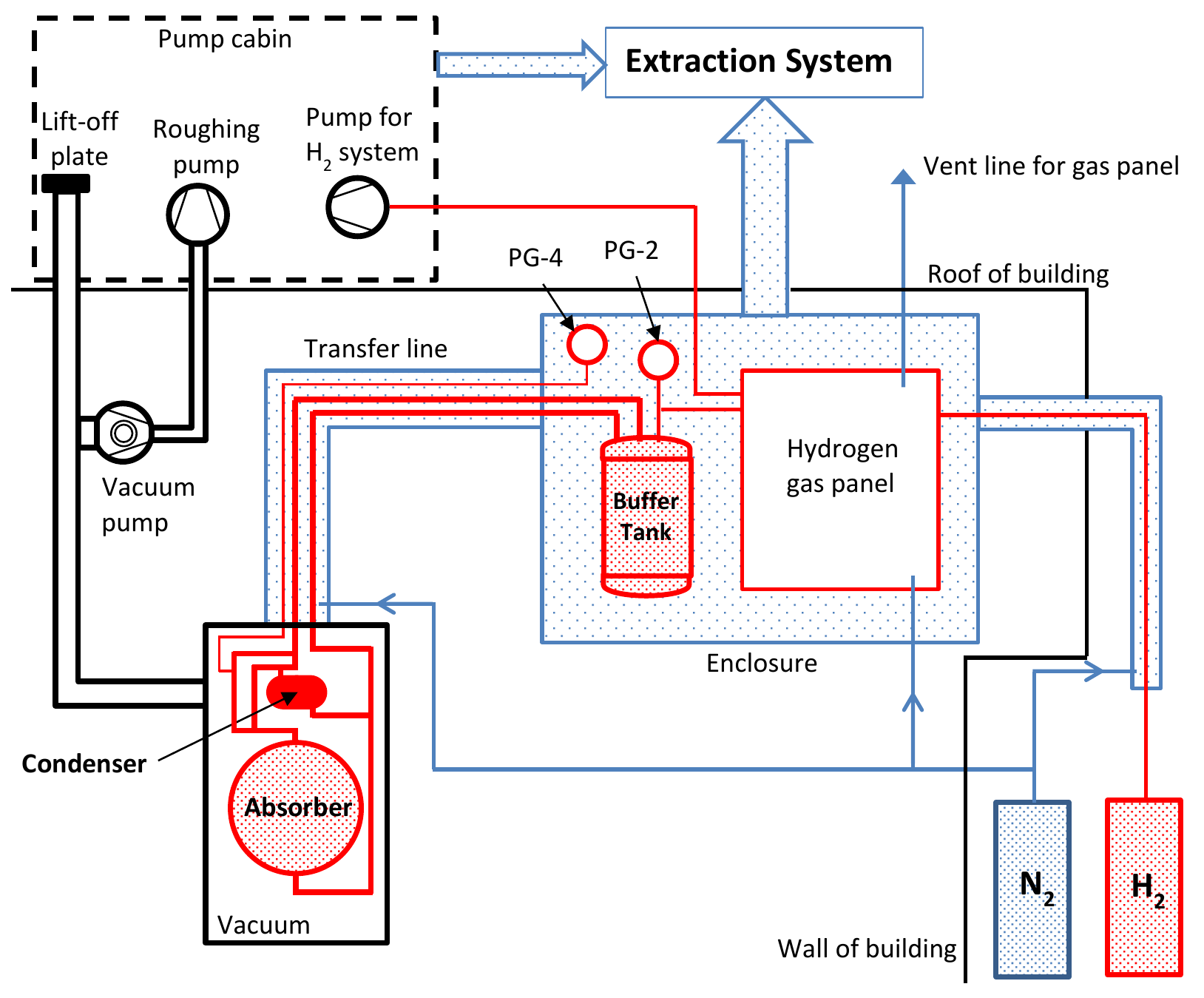}
  \end{center}
  \caption{
    Gas flows in the hydrogen gas system: hydrogen gas is shown in
    red, nitrogen gas is shown in blue. The helium gas used for
    initial flushing of the hydrogen volume, and during cooling, is not shown. 
  }
  \label{Fig:GasPanel:OvrvwGasFlow:Schema}
\end{figure}

A schematic of the flows in the hydrogen gas system is shown in
figure~\ref{Fig:GasPanel:OvrvwGasFlow:Schema}. 
The system comprised a hydrogen-gas panel inside an enclosure, a
source of hydrogen gas, the condensing unit, a pump for evacuating
hydrogen from the pipework, a pump-set to evacuate
the safety vacuum around the condensing unit,
and nitrogen-gas flushing of the secondary containment. 
This system (from the source, through the gas panel, to the connections to
the absorber vessel) was fabricated entirely from stainless-steel components. 
Pipework was welded wherever possible, X-rayed to ATEX rating where required,
and non-welded joints were sealed using metal gaskets.
The pipework was thoroughly tested for leaks using helium gas.

Hydrogen gas was obtained from a multi-cylinder pack (MCP)
situated in a locked cage outside the experimental hall. 
Also inside this cage were the regulator and the pressure-relief valves
for the section of pipework outside the building. 
This MCP contained ample gas, thus obviating the need to 
break and remake connections which would have risked
contamination with air.
It initially contained 200\,bar and, by the end of the
experiment, 170\,bar remained, corresponding to a total consumption
of roughly 20,000\,bar\,\textit{l} of gas.

The hydrogen gas flowed from the regulator in the cage to the gas
panel inside the experimental hall via a jacketed line which was
formed from two concentric tubes. The inner tube carried the hydrogen
gas while the outer tube carried nitrogen gas flowing at a rate of
$\sim$\,1\,\textit{l}/min. to flush out any hydrogen that might leak from
the inner tube.  
This nitrogen gas flowed into the enclosure that surrounds the
gas panel and was then extracted,
via large diameter ($>250$\,mm
bore) pipes, by fans on the roof of the building
and expelled to the atmosphere approximately 4.5\,m
above roof level.

The condensing unit was supplied with room-temperature gas
via a transfer line comprising: 
two 22\,mm bore tubes to convey the gas (initially designed as a
separate feed and return but later used in parallel); one
$\sim$\,4\,mm bore tube which enabled a pressure gauge (PG-04)
situated in the gas-panel enclosure to measure the gas pressure above the
absorber vessel; and a surrounding pipe (110\,mm bore) which formed
the jacket for the flowing nitrogen gas ($\sim$\,1\,\textit{l}/min.) that would
carry any leaked hydrogen gas back into the gas-panel enclosure from
where it would be safely extracted.
The pressure of the gas supplied to the condensing unit was measured
at the outlet of the gas panel (PG-02). 
Any significant difference in pressure between PG-02 and PG-04 would
have revealed a blockage in the pipework at low temperatures. 

There were ten hydrogen sensors, in pairs at five positions:
\begin{itemize}
\item  In the enclosure to detect leakage from the hydrogen-filled
pipework to the nitrogen gas jacket;
\item  In the extraction system for the same reason;
\item  In the pump cabin to detect leakage into the safety vacuum around
the condensing unit or hydrogen pumped from the gas panel;
\item  In the vent line that is purged continuously by nitrogen gas; and
\item  In the experimental hall to detect hydrogen gas leakage from the
secondary containment system.
\end{itemize}

\subsection{Safety volume and quench line}
\label{Subsect:GasPanel:InsVacQL}

The absorber vessel was situated in the bore of the
focus-coil magnet, with thin aluminium
safety windows on both sides forming a safety volume (SV), as shown in 
figure~\ref{Fig:AbsorberVessel:Diag}.
The safety windows were thin to minimise scattering of the muon beam.
Therefore, safe pump-down and venting procedures were followed to ensure
that pressure differentials across all safety and vessel windows
were not large enough to compromise their function.
Three Leybold Ceravac CTR 100
transducers of overlapping ranges measured the vacuum pressure  
in the SV between $\sim10^{-4}$\,mbar and 1000\,mbar.
The SV was evacuated to $\sim10^{-4}$\,mbar before cooling
began, to provide a thermally-insulating environment for the condensing unit, and
was then pumped continuously to extract
any hydrogen that may have leaked from the cold condensing unit. 
The high-vacuum sensor (Penning type: Leybold PTR-225) was switched off
and disconnected from its controller
before hydrogen was allowed into the system to avoid the possibility of high
voltages being in contact with leaked hydrogen. 
Gas pumped from the SV was emitted into a locked and ventilated pump cabin
on the roof of the experimental hall.
The cabin air was extracted by fans and released to the
atmosphere about 4.5\,m above roof level.
All the equipment in this cabin was ATEX-rated. 

The large (150\,mm) bore pipe from the SV to the inlet of the
turbo-molecular pump formed part of the emergency quench line. 
If there had been a sudden rupture of the thin aluminium windows of the
absorber vessel that released a large quantity of hydrogen gas into the SV,
and if the pressure in this line had exceeded
atmospheric pressure, the hydrogen would have escaped through a lift-off
plate situated in the pump cabin and thence been safely extracted to atmosphere.

\subsection{Gas panel}
\label{SubSect:GasPanel:GasPanel}

\begin{figure}
  \begin{center}
    \includegraphics[width=0.85\textwidth]{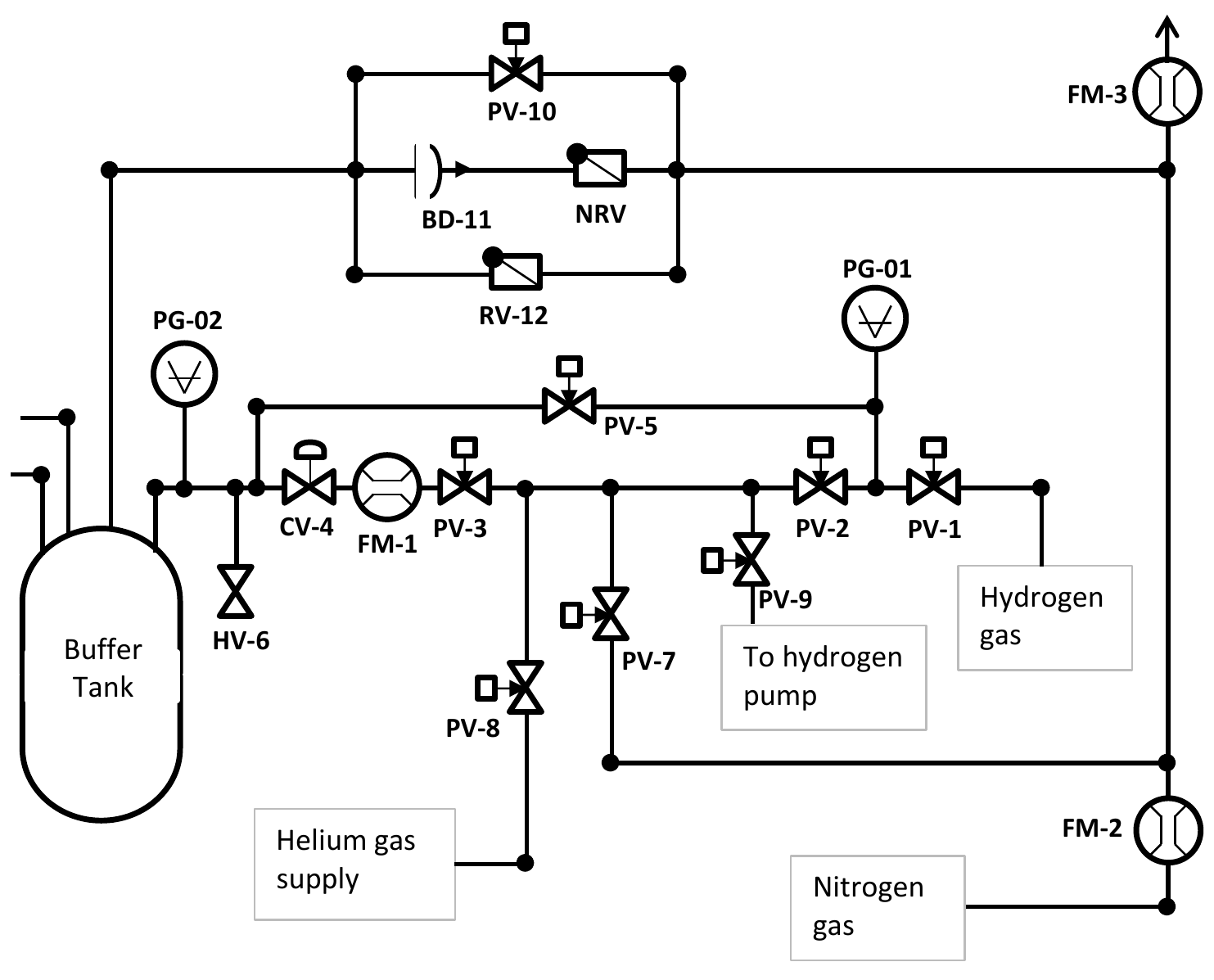}
  \end{center}
  \caption{
    A schematic of the hydrogen gas panel. 
    Key: PG = pressure gauge, CV = control valve, PV = process valve
    (normally closed), HV = hand-operated (manual) valve, RV =
    relief valve, BD = burst disc, NRV = non-return valve,
    and FM = flowmeter. 
    The NRV would prevent back-streaming of nitrogen should the burst
    disc rupture, as this could cause blockages in the low temperature
    pipework. 
    The two tubes leaving the left-hand side of the diagram connect to
    the transfer line. 
  }
  \label{Fig:GasPanel:GasPanel:Proc}
\end{figure}

The gas panel was situated inside an enclosure that contained the nitrogen-gas jacket.
All the valves within the enclosure were actuated by
compressed air which was controlled using electrical actuators
attached to the exterior of the enclosure; the compressed-air tubes
entered the enclosure through grommets.
Pressure sensors were ATEX-rated (IIC) Pepperl and Fuchs 24\,V DC
transducers, PPC-M10, for the range from vacuum to 4\,bar. 
The flow sensor FM-1 was an Ex-Flow Bronkhorst X100 (ATEX-rated IIC).

A schematic of the hydrogen gas panel is shown in
figure~\ref{Fig:GasPanel:GasPanel:Proc}.
Hydrogen gas supplied from the MCP was regulated to about
1350\,mbar, as measured at PG-01; excessive pressures in this
supply line were limited by a 10\,psig relief valve to atmosphere 
outside the building. 
The gas flow (transient rate and cumulative quantity) into the buffer tank,
through the control valve CV-4, was measured by FM-1.
This buffer tank was directly connected (no valves) to the condensing unit via the transfer line.
During cooling and liquefaction, the pressure at PG-02 was automatically
maintained at 1150\,mbar using CV-4. 
The maximum pressure in the buffer tank was limited to 1500\,mbar
by the relief valve, RV-12, which would release excess hydrogen gas
to the vent line that contained
flowing nitrogen gas ($\sim$\,5\,\textit{l}/min.) at atmospheric pressure. 
Also, the process valves PV-7 and PV-10 enabled the operator to release
excess pressure into this vent line. 
A difference in flow rate between flowmeters FM-3 and FM-2 indicated
the amount of gas being released. 
This nitrogen-filled vent line buffered the gas panel outlets (PV-7, PV-10, RV-12, and
BD-11) from air. 
The burst disc was the safety back-up in case the relief
valve failed to open. 
Valve PV-5 bypassed the flow restrictions of FM-1 and CV-4 
to enable the buffer tank and condensing unit to be evacuated
more readily by the hydrogen pump,
and the manual valve HV-6 was used for connecting the leak detector.

\graphicspath{{Figures/}}

\section{Control and monitoring system}
\label{Sect:Cntrl}

The liquid-hydrogen control and monitoring (C\&M) system was sited in
a locked room adjacent to the experimental hall containing MICE.
It was designed to be stand-alone, based on programmable logic controllers
(PLC), and capable of operating without an external
network connection to eliminate the possibility of unauthorised intervention.
A touch-screen display indicated the status of the system
and allowed an operator to run pre-determined sequences such as `Purge', `Fill', and
`Empty'.
The control of individual system components (valves, pumps, compressor etc.) was
also possible using this touch-screen. 
Keyed switches were provided to allow manual override (by authorised persons only) of certain
automatic functions should this be required.
The control PLC was an Omron CJ1M-CPU13-ETN PLC and the touch-screen
an Omron NS8-TV01B-V2.
The control cabinet contained Intrinsically Safe Barriers
(Pepperl and Fuchs), Flowmeter Readout Unit (Bronkhorst), Vacuum Gauge
Controller (Leybold) and the Temperature Monitors (Lakeshore 218). 

\begin{figure}
  \begin{center}
    \includegraphics[width=0.85\textwidth]{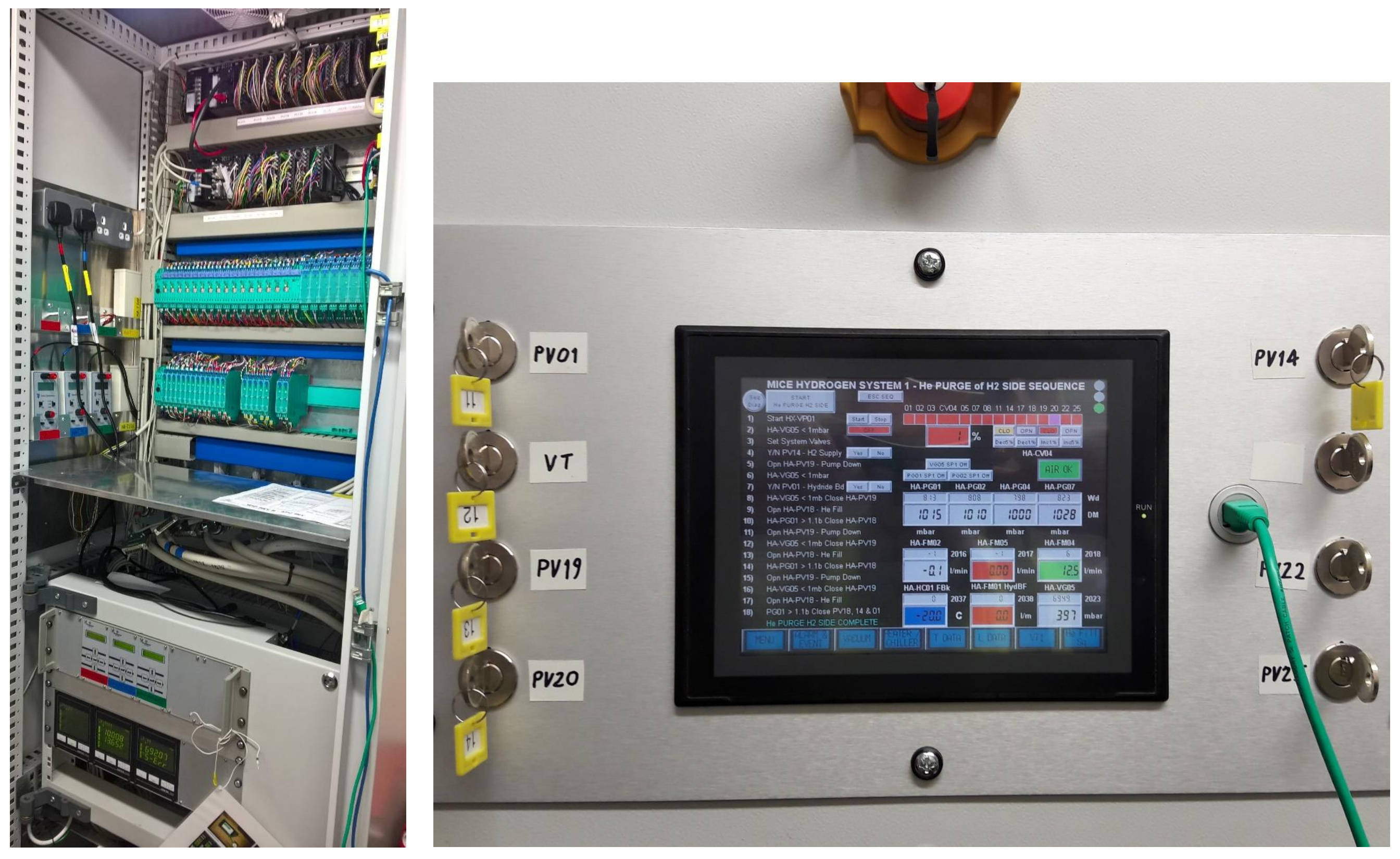}
  \end{center}
  \caption{
    Photographs of the H${_2}$-system control cabinet (left), and the touch-screen panel (right)
between the keyed over-ride switches and below the emergency shut-down button.
  }
\label{Fig:Controls:Diag}
\end{figure}

A second PLC was used for the gas-detection system, which incorporated
an Oliver IGD TOCSIN 920 control panel, and the extraction system.
The extraction system used two 7.5\,kW non-sparking fans in parallel, 
each controlled by an Omron MX2-Inverter.
These fans ensured that the gas-panel enclosure was kept at a
pressure below atmospheric.
One fan alone was capable of maintaining the required air flow; their
speeds were reduced to 50\% when both fans were running.
If escaped hydrogen had been detected, both fans would have automatically run at
100\% to clear all this hydrogen gas as quickly as possible. 
The fans were fitted with differential pressure switches
(RedBin-P500-2) to monitor their operation.
The status of the inverters was also monitored and displayed on the
NS8 touch-screen. 
An uninterruptable power supply (UPS: Rello MST 20-T4-1) of adequate size
allowed the extraction system, in the event of a power failure, the capability
to run the fans for a sufficiently long period during which the entire system
would have been emptied of hydrogen.
The control PLC and gas-detection system were also powered by this
UPS unit.

The `intrinsically-safe' explosion-protection method was used for all
equipment in the gas panel and all of the sensors in the absorber.
Since the heaters attached to the absorber vessel and the coldhead
could not be intrinsically safe, they were interlocked to prevent
their operation unless the vacuum in the SV was better than
$10^{-3}$\,mbar.
Layers Of Protection Analysis (LOPA) was applied to the whole control system
and identified that this interlock required a safety integrity level 1
(SIL1) rating. 
This was achieved by providing two redundant vacuum gauges and gauge
controllers to drive relays to give a dual-guard-line interlock to the
heater power supply.
The LOPA study also identified that a failure of the gas detection
system was a hazard requiring a SIL1 rating. 
To achieve this, the detectors were installed in pairs on separate
wiring loops.

\begin{figure}
  \begin{center}
    \includegraphics[width=0.85\textwidth]{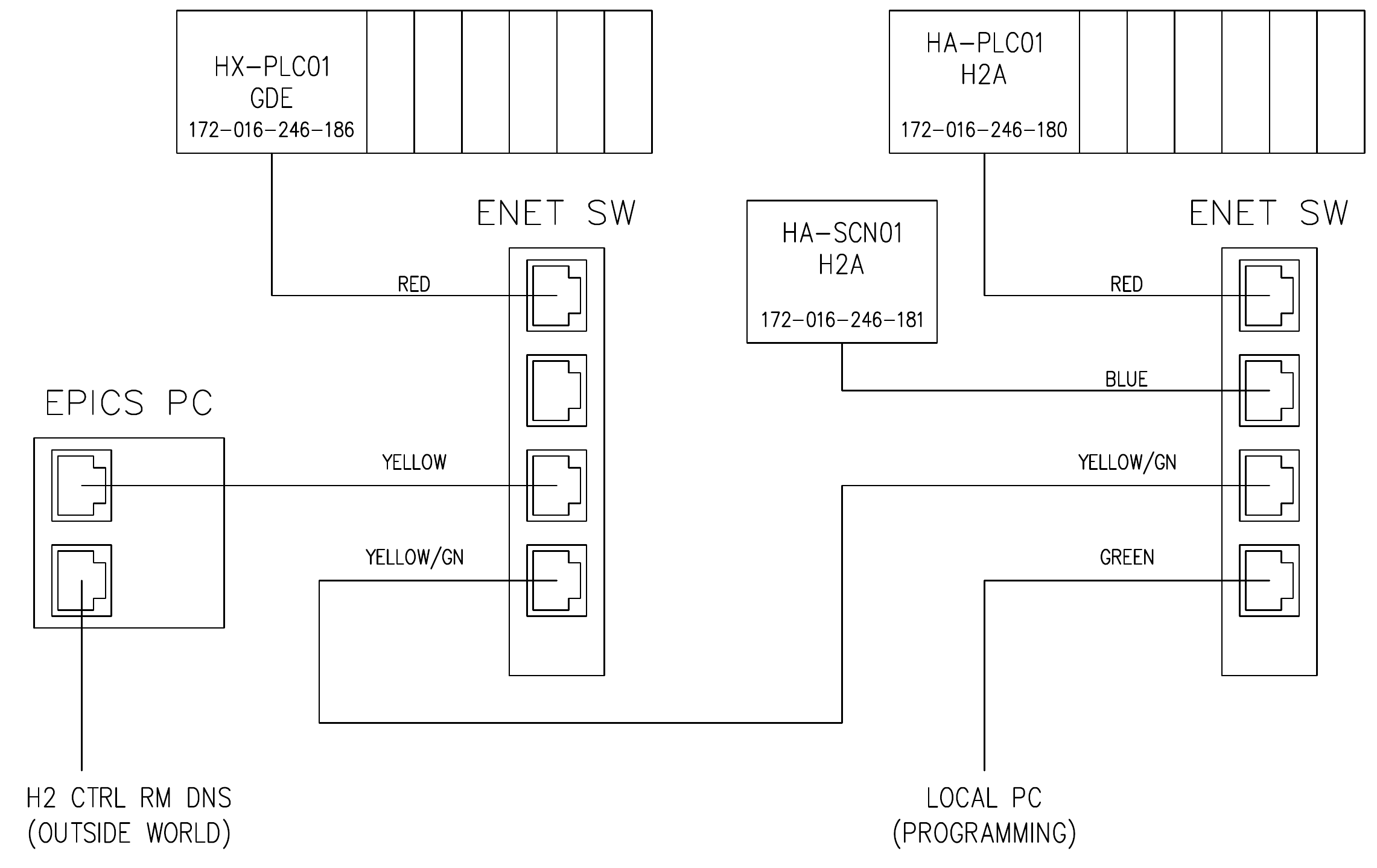}
  \end{center}
  \caption{
    Diagram of the hydrogen-system network and devices: the gas-detection
    (GDE) PLC, and the hydrogen control system (H2A) PLC and touch-screen (SCN).
  }
\label{Fig:Ethernet:Diag}
\end{figure}

The hydrogen system was monitored via a read-only gateway 
on the ethernet network and values were logged
using the Experimental Physics and Industrial
Control System (EPICS). 
Information from the Omron PLCs was retrieved using
the Factory Interface Network Service
(FINS) protocol which was implemented as a module in EPICS.
This provided direct access to the memory registers of the PLCs and
thus allowed the variables used by the PLCs to be provided as EPICS
variables. 
Only the parameters of most interest (e.g. temperatures, pressures and valve
status) were provided as EPICS variables.  
These were then used by the MICE Archiver and Alarm
Handler in the same way as for the other MICE C\&M data. 
To ensure that unauthorised remote operation of the PLCs was not possible,
the hydrogen-system network was isolated from the rest of the
MICE and RAL site networks.  
A PC running EPICS was used to isolate the networks using two network
cards and a network bridge providing read-only access to the EPICS
variables from the MICE network only.
This is shown in figure~\ref{Fig:Ethernet:Diag}.

\graphicspath{{Figures/}}

\section{Performance} 
\label{Sect:Performance}

To verify the performance of the cryogenic systems and to develop and
refine the control sequences, a test setup was established outside the
MICE Hall.
To avoid the need to establish a second hydrogen-safe gas system, the
system-verification test programme was carried out using helium and
neon.
An identical cryostat was used in a different building to:
\begin{itemize}
  \item Gain experience with the experimental setup;
  \item Fine-tune the gas circuitry;
  \item Implement and test any necessary changes to the pipework,
    thermal insulation, sensors, heaters etc.; and
  \item Scale the likely performance to be achieved with hydrogen in MICE.
\end{itemize}

An initial cool-down of the apparatus, using helium gas, was used to determine the base
temperature of the absorber vessel.
The condenser cooled the gas which preferentially passed through the
lower port of the condenser and down through the
thermally-insulated pipework to the bottom of the absorber vessel.
The warmer gas in this absorber vessel preferentially rose through
its top port and continued upwards through more
thermally-insulated pipework to the top port of the condenser. 
These movements of gas within the system established a self-sustaining
circulating gas flow.
As the condenser and absorber vessel cooled, more room-temperature gas
was drawn in from the supply. 
The incoming gas was cooled by the tubes that were heat-sunk to the
thermal shield before it reached the condensing circuit. 

\subsection{Initial tests with helium gas}
\label{Sect:Subsect:Helium}
Using helium gas at a pressure of about 1150\,mbar absolute,
up to 12\,hours were required
for the gas circulation to be established and thus for the efficient cool-down of
the absorber vessel to begin.
The absorber vessel then cooled at a rate of about $-$6\,K/hour.
This steady cooling of the system prevented the build-up of large
thermal gradients that might have led to the development of leaks. 
The condenser reached 22\,K  approximately 53\,hours after the cryocooler was
started, and the absorber vessel reached 39.5\,K.  
Evacuating the gas circuit resulted in a temperature on stage two of the coldhead of
13\,K.
The minimum temperature in the gas-filled condenser of 22\,K 
would have been insufficient to condense hydrogen which has a
boiling point of 20.3\,K at atmospheric 
pressure~\cite{NIST:Chem:WebBook:WWW}. 
This prompted us to improve the cryogenics of the gas circuit.

\begin{table}
\caption{
Temperatures in the condenser and absorber vessel before and after modifications
to the cryogenics.
}
\label{Tab:Perf:Temps}
\begin{center}
\begin{tabular}{|l|c|c|}
\hline
Temperatures (K)  & Before modification & After modification \\
\hline
Condenser With Gas & 22 & 12 \\
Condenser Evacuated & 13 & 4.5 \\
Absorber With Gas & 39.5 & 19.5 \\
\hline
\end{tabular}
\end{center}
\end{table}

\begin{figure}
  \begin{center}
    \includegraphics[width=0.95\textwidth]{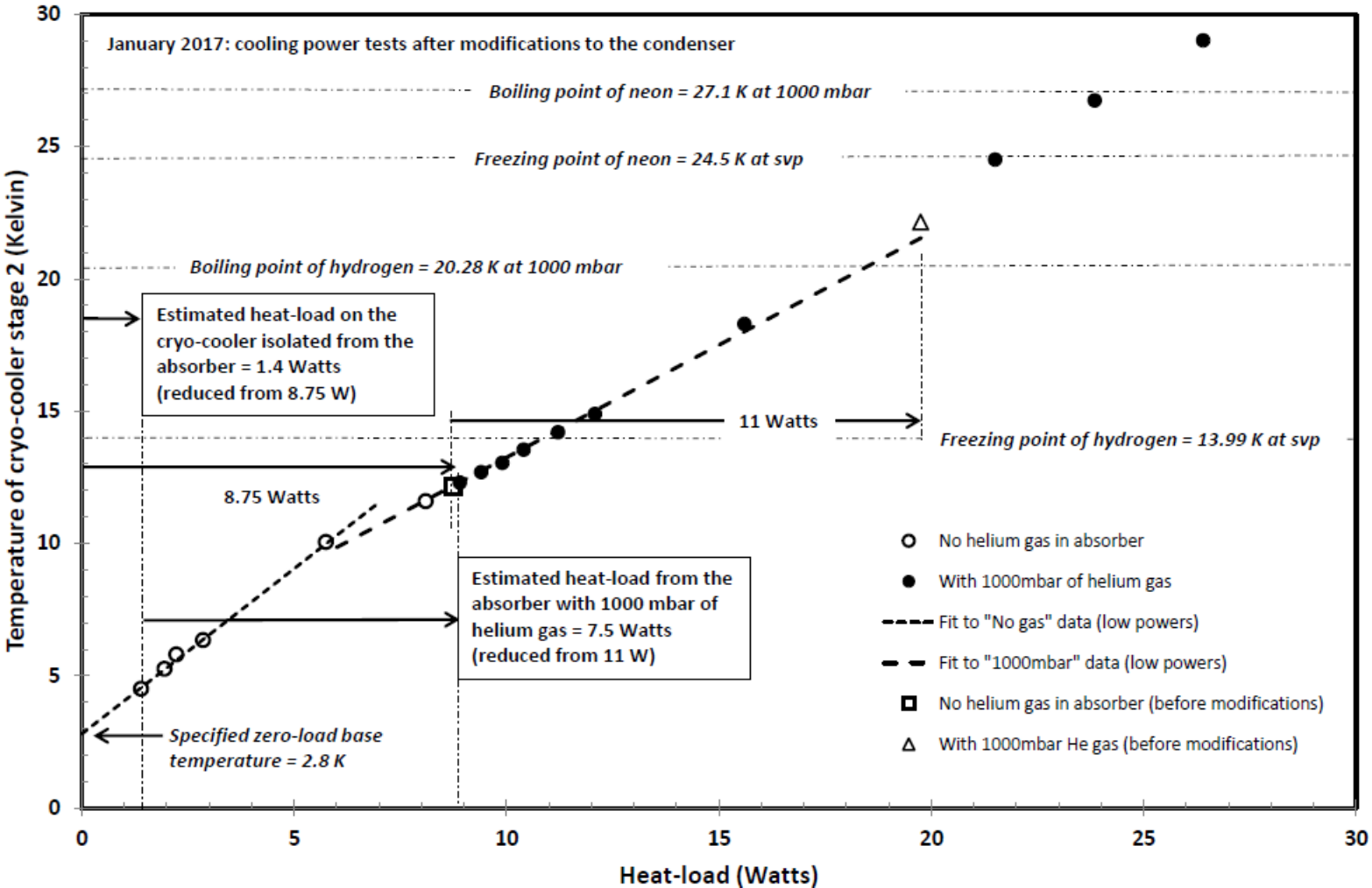}
  \end{center}
  \caption{
    Cryogenic performance of the modified liquefaction circuit.
The temperature of the second stage of the coldhead is plotted as a function
of heat-load (environmental thermal load plus the Joule heating provided
by a heater attached to the second stage). 
The data with the system evacuated was translated horizontally until the
intercept on the temperature axis corresponded to the specified
zero-load temperature of 2.8\,K quoted by the manufacturer of the coldhead. 
This estimates the environmental thermal load on the coldhead and condenser.
Helium gas at a pressure of $\sim$1000\,mbar was then introduced
to the condenser, and this gas cooled the absorber vessel.
The gas-load data were then aligned with the vacuum data
to estimate the environmental thermal load
on the absorber vessel. 
  }
  \label{Fig:Perf:HeatLoad}
\end{figure}

After modifications to the pipework and the thermal insulation, a second run
with helium gas achieved 12\,K on the coldhead and 19.5\,K  in the absorber
vessel. 
And, with the gas circuit evacuated, the temperature of the cold-head
dropped to 4.5\,K. 
This is shown in table~\ref{Tab:Perf:Temps}.
These temperatures gave reassurance that hydrogen could now be liquefied
by the system.
The difference in temperatures, $\Delta T$, between the condenser and the
absorber vessel was also expected to decrease when liquid collected in the absorber vessel.
Figure~\ref{Fig:Perf:HeatLoad} shows the performance of the cryogenic
system before and after the modifications.
The measurements showed that the heat-load on the condenser had been reduced
from 8.75\,W to 1.4\,W by these modifications, and
the additional heat-load from the gas-cooling of the absorber vessel
had been reduced from 11\,W to 7.5\,W. 

\subsection{Liquefaction of neon}

\subsubsection{Off-line}

The next step was to use neon gas as the refrigerant in order to test
the ability of the system to liquefy gas and then maintain a constant
volume of liquid, and to measure the $\Delta T$ between the condenser
and the absorber vessel.
With neon gas flowing into the system, the temperature of the coldhead
ultimately settled at 27.1\,K 
and, once liquid began to collect, the absorber vessel cooled rapidly from 40\,K
to 27.8\,K (the boiling point of neon at $\sim$120\,mbar above atmospheric pressure
~\cite{NIST:Chem:WebBook:WWW}).
With neon being liquefied,
$\Delta T \le$ 1\,K  indicated good circulation of gas and liquid. 
Approximately 2\,\textit{l} of liquid neon were collected 
inside the absorber vessel; both the lowest sensor at the bottom of the vessel 
and the next-lowest sensor
at $45^{\circ}$ from the vertical registered the presence of liquid. 
The heat of the incoming gas during the fill process, and of the
boil-off gas from the absorber vessel, was sufficient to keep the condenser above
the neon freezing temperature of $\sim$24.5\,K. 
After the external supply of gas was stopped, thus isolating
the condenser/absorber circuit, the heater on the coldhead
could control the system pressure above atmospheric and keep the liquid level constant.
This initial test demonstrated that neon gas could be
condensed into the absorber, that the volume of condensed liquid
could be controlled, that freezing of the liquid could be prevented
via the heater on the coldhead, and that $\Delta T$ was small.

\subsubsection{Within MICE}

The cool-down and liquefaction of neon gas was successfully repeated
after the system was installed within the bore of the focus-coil module
in MICE.
As can be seen in table~\ref{Tab:Perf:GasData}, the energy that must be
extracted to cool gas from room temperature and produce the liquid
at its boiling point is less for hydrogen than it is for neon.
This gave the confidence required to progress to liquefying hydrogen.

\subsection{Liquefaction of hydrogen}

\begin{table}
  \caption{
    Some cryogenic data for the gases involved in this experiment. The
expansion ratio is the ratio of the volume of gas to the volume of liquid it creates when condensed.
($T_{0} - T_{bp}$) is the temperature difference between room temperature (295\,K) and the boiling point. The bottom row 
gives the energy that must be extracted from room temperature gas to create the liquid at its boiling point.
  }
  \label{Tab:Perf:GasData}
  \begin{center}
\begin{tabular}{|l|c|c|c|}
 \hline
   & Helium & Hydrogen & Neon \\
\hline
Gas density (kg/m$^{3}$) & 0.16 & 0.08 & 0.82 \\
Expansion ratio & 740 & 830 & 1412 \\
Mass of one liquid litre (kg) & 0.125 & 0.071 & 1.2 \\
Specific heat capacity $c_{p}$ (kJ kg$^{-1}$ K$^{-1}$) & 5.19 & 14.32 & 1.03 \\
$T_{0} - T_{bp}$ (K) & 291 & 275 & 268 \\
Latent Heat (kJ/kg) & 21 & 455 & 86.3 \\
Energy per liquid litre (kJ) & 191 & 312 & 435 \\
\hline
\end{tabular} 
\end{center}
\end{table}

The absorber vessel was pre-cooled using helium gas
to transport heat between the condenser and the absorber vessel. 
This pre-cool lasted for four days, and ended with the following
temperatures:
\begin{itemize}
  \item Coldhead first stage: 44.1\,K;
  \item Coldhead second stage: 14.0\,K;
  \item Absorber top: 21.2\,K; and
  \item Absorber bottom: 20.4\,K.
\end{itemize}

\begin{figure}
  \begin{center}
    \includegraphics[width=0.95\textwidth]{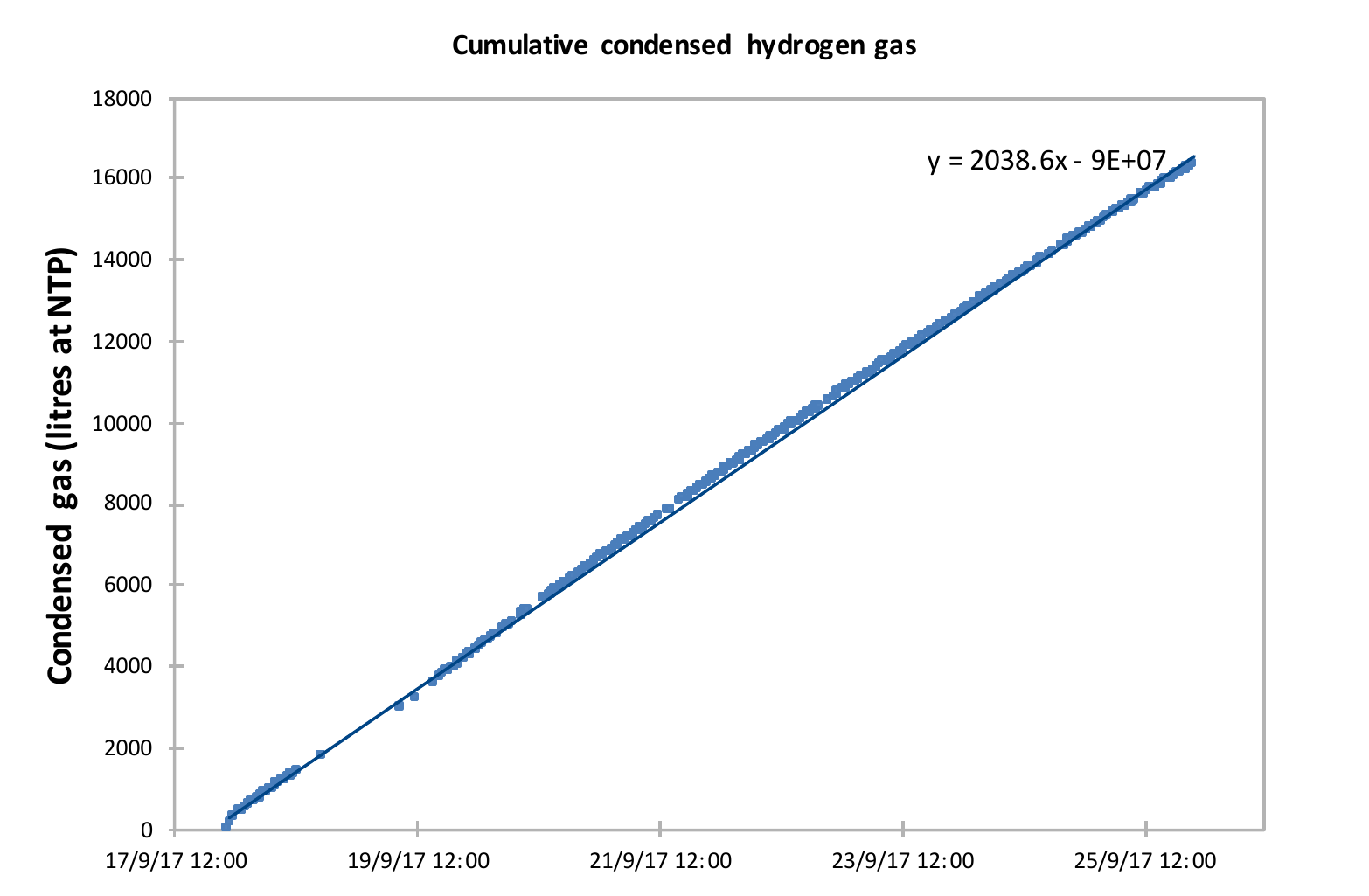}
  \end{center}
  \caption{
    Filling the absorber vessel with condensing hydrogen gas. The cumulative quantity 
of condensed gas increased almost linearly over about 8\,days, at an average
rate of just over 2000\,bar litres per day.
  }
  \label{Fig:Perf:CoolDown}
\end{figure}

After emptying the system of helium gas, hydrogen gas at 1150\,mbar was introduced.
The incoming gas warmed the condenser, and the automatically-controlled
heater maintained the coldhead second stage temperature above 18\,K.
If this temperature had decreased to below 15\,K, the control system would have
automatically switched off the helium compressor that powered the coldhead
to prevent freezing of the hydrogen.
Final cool-down and liquefaction of hydrogen took eight days.
The liquefaction rate (as measured by gas passing through FM-1) was almost uniform, 
as shown in figure~\ref{Fig:Perf:CoolDown}. 
A total of $\sim$16,300\,bar \textit{l} of hydrogen gas were liquefied.

\begin{figure}
  \begin{center}
    \includegraphics[width=0.95\textwidth]{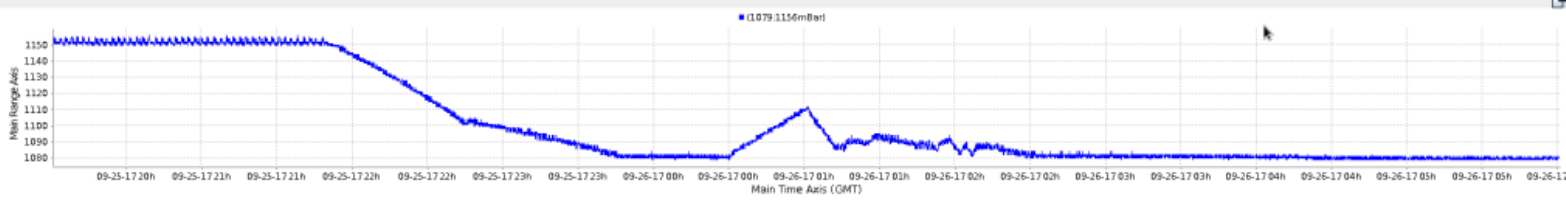}
  \end{center}
  \caption{
    Pressure in the absorber vessel, as a function of time, during the change-over from the
    filling process (p = 1150\,mbar) to the pressure-controlling state
    (1080\,mbar).
  }
  \label{Fig:Perf:Pressure}
\end{figure}

Once the absorber vessel was deemed to be full (as indicated by the
volume of gas condensed and the level sensors) the control sequence
was changed from `Fill' to `Full' and the external supply of hydrogen
was isolated. The operating pressure then decreased to the new setting
of 1080\,mbar.
Figure~\ref{Fig:Perf:Pressure} shows the pressure in the absorber
vessel during this change-over. 
The control system then reliably maintained the pressure around this value.
If the pressure had decreased below 1040\,mbar, the helium compressor
would have been switched off automatically.

The 22\,\textit{l} volume of liquid hydrogen was maintained for the
duration of this phase of the MICE data-taking from the
25$^{\rm th}$ September 2017 to the 16$^{\rm th}$ October 2017.
The coldhead was then switched off and the heaters were switched on,
delivering a nominal power of 50\,W to the absorber vessel. 
The hydrogen gas vented to the atmosphere via RV-12 and the
nitrogen-purged vent line.
Emptying the absorber vessel took approximately 5.5\,hours. 
The hydrogen gas was then purged from the system.

\graphicspath{{08-Conclusions/Figures/}}

\section{Conclusions}
\label{Sect:Conclusions}

A complete system capable of safely
condensing hydrogen gas in a vessel with thin aluminium windows
was designed, constructed and operated. 
This vessel was placed inside the focus-coil magnet of the MICE experiment at the
Rutherford Appleton Laboratory and was irradiated with a beam of muons.
Approximately 22~\textit{l} of liquid hydrogen were collected in this vessel,
and this liquid was kept at a constant temperature and pressure for three weeks.
This enabled the MICE collaboration to measure the loss of energy
and change of trajectory of muons in liquid hydrogen,
thus elucidating the details of the interactions that lead to beam-cooling effects in liquid hydrogen.
\section*{Acknowledgements}

The work described here was performed to deliver the liquid-hydrogen
absorber system for the international Muon Ionization Cooling
Experiment built at the STFC Rutherford Appleton Laboratory (RAL) in
the UK.
We are indebted to the MICE collaboration for providing the motivation
for, and the context within which, the work reported here was carried
out.
We would like to acknowledge the support and hospitality of FNAL, The
Daresbury Laboratory, KEK, RAL, and the Universities of Mississippi and
Oxford where designs, machining or test procedures were carried out.
The work described here was made possible by grants from the
Department of Energy and the National Science Foundation (USA), the
Science and Technology Facilities Council (UK) and the Japan Society
for the Promotion of Science.
We gratefully acknowledge all sources of support.
We are grateful for the support given to us by the staff of the STFC
Rutherford Appleton and Daresbury Laboratories during the build,
commissioning and operational phases of the project.

\end{document}